\PassOptionsToPackage{fleqn}{amsmath}
\documentclass{sig-alternate}
\usepackage{mathptmx} 

\newcommand{\ignore}[1]{}
\usepackage[pass]{geometry}

\usepackage{fancyhdr}
\usepackage[normalem]{ulem}
\usepackage[hyphens]{url}
\usepackage[sort,nocompress]{cite}
\usepackage[final]{microtype}
\usepackage{flushend}
\usepackage{comment}
\usepackage{amsmath}
\usepackage{mathtools}
\usepackage{pifont}
\usepackage{wrapfig,lipsum,booktabs}
\usepackage{multirow}
\usepackage{color}
\usepackage[bookmarks=true,breaklinks=true,letterpaper=true,colorlinks,linkcolor=black,citecolor=blue,urlcolor=black]{hyperref}

\usepackage{mathtools}

\DeclarePairedDelimiter\floor{\lfloor}{\rfloor}
\usepackage{amsmath}

\usepackage{listings, color, setspace, graphics}
\usepackage{fancyhdr, amsmath, amssymb, setspace, amssymb}
\usepackage{xspace}

\newcommand{\scheme}{\textsc{GraphR}}
\newcommand{\onurgraph}{\textsc{Tesseract}}

\definecolor{orange}{rgb}{1,0.5,0}

\usepackage{pifont}
\usepackage{graphicx}
\lstdefinelanguage{scala}{
  morekeywords={abstract,case,catch,class,def,%
    do,else,extends,false,final,finally,%
    for,if,implicit,import,match,mixin,%
    new,null,object,override,package,%
    private,protected,requires,return,sealed,%
    super,this,throw,trait,true,try,%
    type,val,var,while,with,yield,%
    processEdge,reduce,apply,%
    local, put, get,end,
    Send, Recv, Sync, SendBatch, RecvBatch, InitBatch},
  otherkeywords={=>,<-,<\%,<:,>:,\#,@},
  sensitive=true,
  morecomment=[l]{//},
  morecomment=[n]{/*}{*/},
  morestring=[b]",
  morestring=[b]',
  morestring=[b]"""
}

\definecolor{dkgreen}{rgb}{0,0.6,0}
\definecolor{gray}{rgb}{0.5,0.5,0.5}
\definecolor{mauve}{rgb}{0.58,0,0.82}

\usepackage{inconsolata}
\lstdefinestyle{MyScala}{
  numbers=left,  
  firstnumber=1,
  numberfirstline=true,
  numberstyle=\scriptsize\ttfamily,
  language=scala,
  aboveskip=3mm,
  belowskip=3mm,
  showstringspaces=false,
  basicstyle={\small\ttfamily},
  keywordstyle=\bfseries,
  captionpos=b,
  columns=flexible,
  xleftmargin=0.05\textwidth, xrightmargin=0\textwidth,
  breaklines=true,
  breakatwhitespace=true,
  mathescape=true,
  tabsize=4
}

\lstdefinestyle{MyText}{
  language=scala,
  aboveskip=3mm,
  belowskip=3mm,
  showstringspaces=false,
  basicstyle={\scriptsize\ttfamily},
  keywordstyle=\bfseries,
  captionpos=b,
  columns=flexible,
  xleftmargin=.09\textwidth, xrightmargin=.09\textwidth,
  breaklines=true,
  breakatwhitespace=true,
  mathescape=true,
  tabsize=4
}

\newcommand{\microsubmissionnumber}{008}

\fancypagestyle{firstpage}{
  \fancyhf{}
\setlength{\headheight}{50pt}

  \fancyhead[C]{\normalsize{HPCA 2018 Submission
      \textbf{\#\microsubmissionnumber} -- Confidential Draft -- Do NOT Distribute!!}}
  \pagenumbering{arabic}
}

\title{GraphR: Accelerating Graph Processing Using ReRAM }

\usepackage{kantlipsum}
\makeatletter
\def\@maketitle{\newpage
 \null
 \setbox\@acmtitlebox\vbox{%
\baselineskip 20pt
\vskip 0cm                  
   \begin{center}
   \vskip -0.5cm 
    {\ttlfnt \@title\par}       
    \vskip 0em                
{\subttlfnt \the\subtitletext\par}\vskip 1.25em
    {\baselineskip 16pt\aufnt   
     \lineskip 0em             
     \begin{tabular}[t]{c}\@author
     \end{tabular}\par}
    \vskip 2em               
   \end{center}}
 \dimen0=\ht\@acmtitlebox
 \unvbox\@acmtitlebox
 \ifdim\dimen0<0.0pt\relax\vskip-\dimen0\fi}
\makeatother

\newcommand\Mark[1]{\textsuperscript#1}

\begin{document}

\author{%
Linghao Song\Mark{*}, Youwei Zhuo\Mark{\#}, Xuehai Qian\Mark{\#}, Hai Li\Mark{*} and Yiran Chen\Mark{*}\\
\Mark{*}Duke University, \Mark{\#}University of Southern California\\
\email{\Mark{*}\{linghao.song, hai.li, yiran.chen\}@duke.edu, 
\Mark{\#}\{youweizh, xuehai.qian\}@usc.edu
}
}

\maketitle
\pagestyle{plain}

\let\thefootnote\relax\footnotetext{A version submitted to \url{https://arxiv.org}. This paper is to appear in HPCA 2018.}


\begin{abstract}
Graph processing recently received intensive interests 
in light of a wide range of needs to understand relationships.
It is well-known for the {\em poor locality} 
and {\em high memory bandwidth requirement}.
In conventional architectures, they incur a significant amount of data movements 
and energy consumption
which motivates several hardware graph processing accelerators.
The current graph processing accelerators rely on memory access optimizations 
or placing computation logics close to memory.
Distinct from all existing approaches, we leverage an emerging memory 
technology to accelerate graph processing with analog computation. 

This paper presents \scheme, the first ReRAM-based graph processing 
accelerator. \scheme\ follows the principle of near-data processing
and explores the opportunity of performing massive parallel analog 
operations with low hardware and energy cost.
The analog computation is suitable for graph processing because:
{\em 1)} The algorithms are iterative and could inherently
tolerate the imprecision;
{\em 2)} Both probability calculation (e.g., PageRank and Collaborative Filtering) and
typical graph algorithms involving integers (e.g., BFS/SSSP) are resilient to errors. 
The key insight of \scheme\ is that if a vertex program of a graph algorithm
can be expressed in sparse matrix vector multiplication (SpMV), it can be 
efficiently performed by ReRAM crossbar. We show that this assumption 
is generally true for a large set of graph algorithms. 

\scheme\ is a novel accelerator architecture consisting of two components:
{\em memory ReRAM} and {\em graph engine (GE)}. 
The core graph computations are performed in sparse matrix format in GEs (ReRAM crossbars).
The vector/matrix-based graph computation is not new, but 
ReRAM offers the unique opportunity to realize the massive parallelism with 
unprecedented energy efficiency and low hardware cost.
With small subgraphs processed by GEs,
the gain of performing parallel operations overshadows
the wastes due to sparsity.
The experiment results show that \scheme\ achieves a 16.01$\times$ (up to 132.67$\times$) speedup and a 33.82$\times$ energy saving on geometric mean compared to a CPU baseline system.
Compared to GPU,
\scheme\ achieves 1.69$\times$ to 2.19$\times$ speedup and
consumes 4.77$\times$ to 8.91$\times$ less energy.
\scheme\ gains a speedup of 1.16$\times$ to 4.12$\times$, and 
is 3.67$\times$ to 10.96$\times$ more energy efficiency compared to PIM-based architecture.




\label{_abstract}
\end{abstract}

\section{Introduction}
\label{_introduction}
With the explosion of data collected from massive sources, 
graph processing received intensive interests 
due to the increasing needs to understand relationships.
It has been applied in many important domains including
cyber security~\cite{vigna1998netstat},
social media~\cite{agichtein2008finding},
PageRank citation ranking~\cite{page1999pagerank},
natural language processing~\cite{ biemann2006chinese, corston2005system, mihalcea2011graph}, system biology~\cite{chesler2005complex, conesa2005blast2go},  recommendation systems~\cite{linden2003amazon,schafer2007collaborative,walter2008model}
and machine learning~\cite{frey1998graphical, wainwright2008graphical, murphy2012machine}.


There are several ways to perform graph processing. 
The distributed systems~\cite{low2012distributed, gonzalez2012powergraph, xin2013graphx, chen2015powerlyra, 7832830, 7573805} leverage the ample computing resources to process
large graphs.
However, they inherently suffer from 
synchronization and fault tolerance overhead~\cite{khayyat2013mizan, randles2010comparative, zhao2014lightgraph} and load imbalance~\cite{wang2014replication}.
Alternatively, disk-based single-machine graph processing systems
~\cite{zhu2015gridgraph, kyrola2012graphchi, girod2008xstream, vora2016load, 7840840}
(a.k.a. out-of-core systems) can largely
eliminate all the challenges of distributed frameworks.
The key principle of such systems is 
to keep only a small portion of active graph data in memory and spill 
the remainder to disks.
The third approach is the in-memory graph processing.
The potential of in-memory data processing has been exemplified in
a number of successful projects, including RAMCloud~\cite{ongaro2011fast}, 
Pregel~\cite{malewicz2010pregel}, GraphLab~\cite{low2014graphlab}, Oracle TimesTen~\cite{lahiri2013oracle}, 
and SAP HANA~\cite{farber2012sap}.



It is well-known for the {\em poor locality} because of the random accesses in traversing the 
neighborhood vertices, and {\em high memory bandwidth requirement}, 
because the computations on data accesses from memory are typically simple.
In addition, graph operations lead to memory bandwidth waste because
they only use a small 
portion of a cache block. 
In conventional architecture, 
 graph processing incurs significant amount of 
data movements and energy consumption. 
To overcome these challenges, 
several hardware accelerators~\cite{ahn2015scalable,ozdal2016energy,hamgraphicionado} 
are proposed to execute graph processing more efficiently
with specialized architecture. 
In the following, we retrospect the graph computation and the current solutions 
to motivate our approach.

A graph can be naturally represented as an adjacency matrix
and most graph algorithms can be implemented by 
some form of matrix-vector multiplications.
However, due to the sparsity of graph, 
graph data are not stored in compressed sparse matrix representations,
instead of matrix form.
Graph processing based on sparse data representation involves: 
{\em 1)} bringing data for computation from memory based on compressed representation;
{\em 2)} performing the computations on the loaded data. 
Due to the sparsity, the data accesses in {\em 1)} may be {\em random and irregular}.
In essence, {\em 2)} performs {\em simple} computations that are part of the matrix-vector multiplications
but only on non-zero operands.
As a result, each computing core experiences alternative long random memory access latency 
and short computations.
This leads to the well-known challenges in graph processing and 
other issues such as memory bandwidth waste~\cite{hamgraphicionado}. 

The current graph processing accelerators mainly optimize the memory accesses. 
Specifically, Graphicionado~\cite{hamgraphicionado} reduces memory access latency and improves throughput 
by replacing random accesses to conventional memory hierarchy with 
sequential accesses to scratchpad memory optimization and pipelining.
Ozdal {\em et al.}~\cite{ozdal2016energy} improves the performance and energy efficiency by 
latency tolerance and hardware supports
for dependence tracking and consistency.
\onurgraph~\cite{ahn2015scalable} applies the principle of near-data processing 
by placing compute logics (e.g., in-order cores)
close to memory to claim the high internal bandwidth of Hybrid Memory Cube (HMC)~\cite{pawlowski2011hybrid}.
However, all architectures do little change on compute unit, --- the simple computations are
performed one at a time by instructions or specialized units.

To perform matrix-vector multiplications, two approaches exist
that reflect two ends of the spectrum:
{\em 1)} the dense-matrix-based methods incur
regular memory accesses and perform computations with every element in matrix/vector;
{\em 2)} the sparse-matrix-based methods
incur random memory accesses but only perform computations on non-zero operands. 
In this paper, we adopt an approach that can be considered as the 
{\em mid-point} between these two ends that could potentially 
achieve better performance and energy efficiency. 
Specifically, we propose to perform sparse matrix-vector multiplications
on data blocks of compressed representation. 
The benefit is two-fold. First, the computation and data movement ratio is increased. 
It means that the cost of bringing data 
could be naturally hidden by the larger amount of computations on a block of data.
Second, inside this data block, computations 
could be performed in parallel.
The downside is that certain hardware and energy will be wasted in 
performing useless multiplications with zero. 

This approach could in principle be applied to the current GPUs or accelerators,
but with the same amount of compute resources (e.g., SM in GPUs), it is unclear 
whether the gain would outweigh the inefficiency caused by the sparsity.
Clearly, a key factor is the cost of compute logic, ---
with a low-cost mechanism to implement matrix-vector multiplications, 
the proposed approach is likely to be beneficial.

In this paper, we demonstrate that 
the non-volatile memory, metal-oxide resistive random access memory (ReRAM)~\cite{wong2012metal}
could serve as the essential hardware building block to perform
matrix-vector multiplications in graph processing. 
Recent works~\cite{shafiee2016isaac, chi2016prime,pipelayer} 
demonstrate the promising applications of efficient 
in-situ matrix-vector multiplication of ReRAM on 
neural network acceleration.
The analog computation is suitable for graph processing because:
{\em 1)} The iterative algorithms could {\em tolerate the imprecise values by nature};
{\em 2)} Both probability calculation (e.g., PageRank and Collaborative Filtering) and
typical graph algorithms involving integers (e.g., BFS/SSSP) are {\em resilient to errors}.  
Due to the low-cost and energy efficiency, 
matrix-based computation in ReRAM would not incur
significant hardware waste due to sparsity.
Such waste is only incurred 
inside the ReRAM crossbar with moderate size (e.g., 8 $\times$ 8).
Moreover, the sparsity is not completely lost, ---
if a small subgraph contains all zeros, it does not need to be processed.
As a result, the architecture will mostly enjoy the benefits
of more parallelism in computation and higher ratio between 
computation and data movements.
Performing computation in ReRAM also 
enables near data processing with reduced data movements: 
the data do not need to go through the memory hierarchy like 
in the conventional architecture or some accelerators.


Applying ReRAM in graph processing poses a few challenges:
{\em 1) Data representation}. Graph is stored in compressed
format, not in adjacency matrix, to perform in-memory computation, 
data needs to be converted to matrix format. 
{\em 2) Large graph}. The real-world large graphs may not fit in memory. 
{\em 3) Execution model}. The order of subgraph processing needs to be carefully determined because it affects the hardware cost and correctness. 
{\em 4) Algorithm mapping}. 
It is important to map various graph algorithms to ReRAM with good
parallelism.

This paper proposes, 
\scheme, a novel ReRAM-based accelerator for graph processing. 
It consists of two key components: memory ReRAM and graph engine (GE), 
which are both based on ReRAM but with different functionality.
The memory ReRAM {\em stores} the graph data in {\em compressed sparse representation}.
GEs (ReRAM crossbars) perform the efficient
{\em matrix-vector multiplications} on the {\em sparse matrix representation}. 
We propose a novel {\em streaming-apply} model and the 
corresponding preprocessing algorithm to ensure correct processing
order. 
We propose two algorithm mapping patterns, 
parallel MAC and parallel add-op, to achieve good parallelism
for different type of algorithms.
\scheme\ can be used as a drop-in accelerator for out-of-core 
graph processing systems.

In the evaluation, we compare \scheme\ 
with a software framework~\cite{zhu2015gridgraph} on a high-end CPU-based platform 
,GPU and PIM~\cite{ahn2015scalable} implementations.
The experiment results show that \scheme\ achieves a 16.01$\times$ (up to 132.67$\times$) speedup and a 33.82$\times$ energy saving on geometric mean compared to the CPU baseline.
Compared to GPU,
\scheme\ achieves 1.69$\times$ to 2.19$\times$ speedup and
consumes 4.77$\times$ to 8.91$\times$ less energy.
\scheme\ gains a speedup of 1.16$\times$ to 4.12$\times$, and 
is 3.67$\times$ to 10.96$\times$ more energy efficiency compared to PIM-based architecture.

This paper is organized as follows. Section~\ref{_background} introduces the 
background of graph processing, ReRAM and current hardware graph processing accelerators.
Section~\ref{graphr_arch} describes \scheme\ architecture.
Section~\ref{graph_engine} presents processing patterns for various graph algorithms.  
Section~\ref{_evaluation} presents the evaluation methodology and experiment results. 
Section~\ref{_conclusion} concludes the paper.



\section{Background and Motivation}
\label{_background}
\subsection{Graph Processing}
\label{_label_graph_processing}

\begin{figure}[h]
\centering
\vspace{-5pt}
\includegraphics[width=0.98\columnwidth]{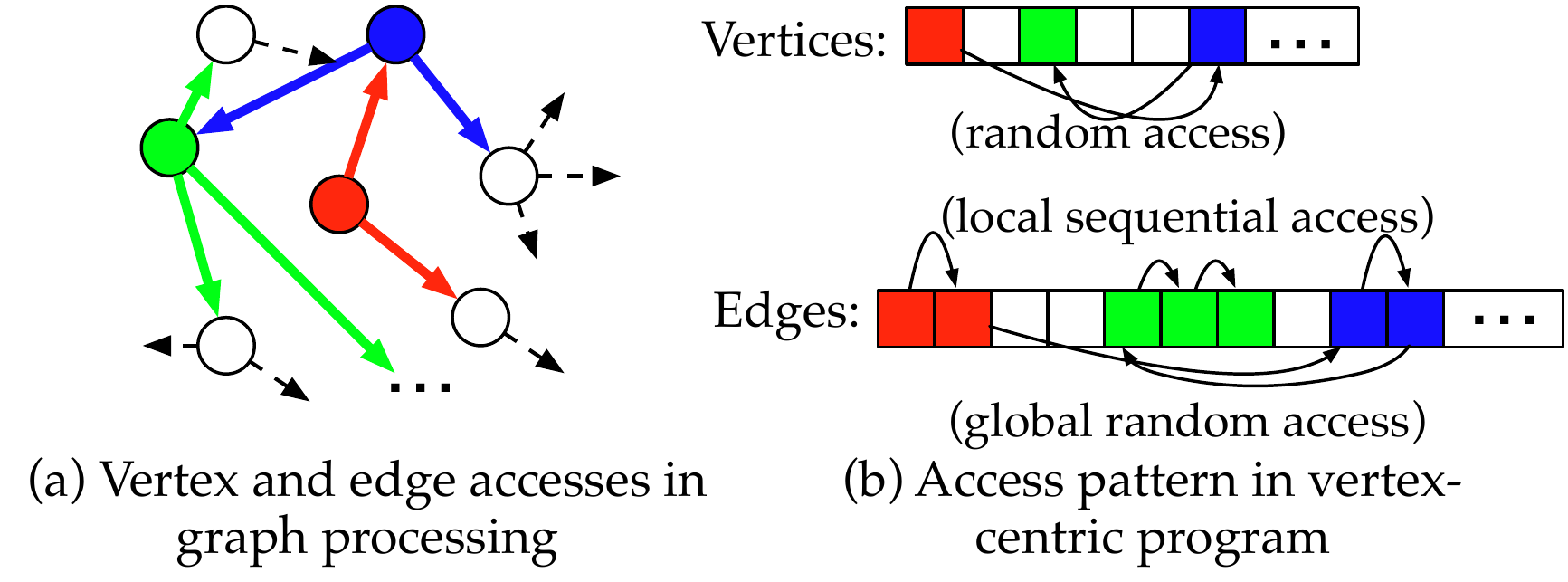}
\vspace{-8pt}
\caption{Graph Processing in Vertex-Centric Program }
\label{vertex_centric}
\vspace{-8pt}
\end{figure}

\begin{figure*}[htb]
\centering
\vspace{-0pt}
\includegraphics[width=1.8\columnwidth]{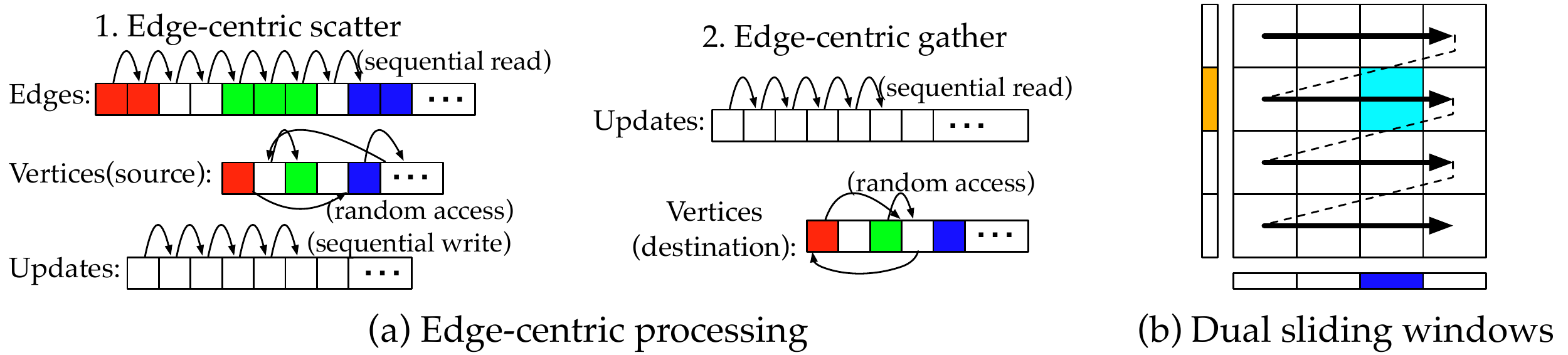}
\vspace{-8pt}
\caption{(a) Edge-Centric Processing and (b) Dual Sliding Windows}
\label{edge_centric}
\vspace{-8pt}
\end{figure*}

Graph algorithms traverse vertices and edges to discover
interesting graph properties based on relationship. 
A graph could be naturally considered as an adjacency matrix, 
where the rows correspond to the vertices and the 
matrix elements represent the edges. 
Most graph algorithms can be mapped to matrix operations.
However, in reality, the graph is {\em sparse}, which 
means that there would be many zeros in the adjacency matrix.
This property incurs the waste of both storage and compute resources. 
Therefore, the current graph processing systems 
use the format that is suitable for sparse graph data. 
Based on such data structures, the graph processing can be essentially considered
as implementing the matrix operations on the sparse matrix representation. 
In this case, individual (and simple) operations
in the whole matrix computation are performed by the compute units 
(e.g., a core in CPU or an SM in GPU) after data 
is fetched. In the following, we elaborate the challenge
of random accesses in various graph processing approaches. 
More details on sparse graph data representation will be discussed 
in Section~\ref{graph_format}.    

To provide an easy programming interface, 
the vertex-centric program featuring 
``Think Like a Vertex (TLAV)''~\cite{malewicz2010pregel} was proposed as
a natural and intuitive 
way for human brain to think of the graph problems. 
Figure~\ref{vertex_centric} (a) shows an example, 
an algorithm could first access and process the 
\textcolor{red}{red} vertex in the center with 
all its neighbors through the \textcolor{red}{red} edges. 
Then it can move to one of the neighbors, 
the \textcolor{blue}{blue} vertex on the top, accessing the vertex and the 
neighbors through the \textcolor{blue}{blue} edges. 
After that, the algorithm can access another vertex 
(the \textcolor{green}{green} one on the left),
which is one of the neighbors of the \textcolor{blue}{blue} vertex. 

In graph processing, the vertex accesses lead
to random accesses because the neighbor vertices 
are not stored continuously.
For edges, they incur {\em local sequential access}
because the edges related to a vertex are stored continuously but
{\em global random accesses} because algorithm needs
to access the edges of different vertices. 
The concepts are shown in Figure~\ref{vertex_centric} (b).
The random accesses lead to long memory latency and, more importantly, 
the bandwidth waste, because only a small portion of data are accessed
in a memory block.
In a conventional hierarchical memory system, this leads to 
the significant data movements.



Clearly, reducing random accesses is critical to improve the 
performance of graph processing, this is particularly crucial for the 
disk-based single machine graph processing systems 
(a.k.a out-of-core systems~\cite{ zhu2015gridgraph, kyrola2012graphchi, girod2008xstream, vora2016load}), 
because the random disk I/O operations are much more detrimental 
to the performance. 
In this context, the memory is considered small and fast (therefore can 
afford random accesses) while 
disk is considered large and slow (therefore should be only accessed sequentially). 
The edge-centric model in X-Stream~\cite{girod2008xstream} is a notable 
solution for reducing random accesses.
Specifically, the edges of a graph partition are stored and accessed
sequentially in disk and the related vertices are stored and 
accessed randomly in memory.
Such setting is feasible because typically the vertex data are much smaller
than edge data. 
During process, X-Stream generates {\em Updates} in {\em scatter} phase, 
which incurs sequential writes, and then, 
applies these Updates to vertices in {\em gather} phase, 
which incurs sequential reads. 
The concepts are shown in Figure~\ref{edge_centric} (a).



A notable drawback of X-stream is that the number of updates may be as large as that of edges, GridGraph~\cite{zhu2015gridgraph} proposed optimizations to further 
reduce the storage cost and data movements due to updates. 
The solution is based on the dual sliding windows (shown
in Figure~\ref{edge_centric} (b)), which
partitions edges into blocks and vertices into chunks. 
On visiting the edge blocks,
the source vertex chunk 
(\textcolor{orange}{orange}) is accessed and updates are 
directly applied to the destination vertex chunk (\textcolor{blue}{blue}).
This requires no temporary update storage as in X-Stream. 
Edge blocks can be accessed in source oriented order 
(shown in Figure~\ref{edge_centric} (b)) or destination oriented order. 
Note that the dual sliding window mechanism is based on edge-centric model.



\subsection{ReRAM Basics}

\begin{figure}[h]
\centering
\vspace{-5pt}
\includegraphics[width=0.9\columnwidth]{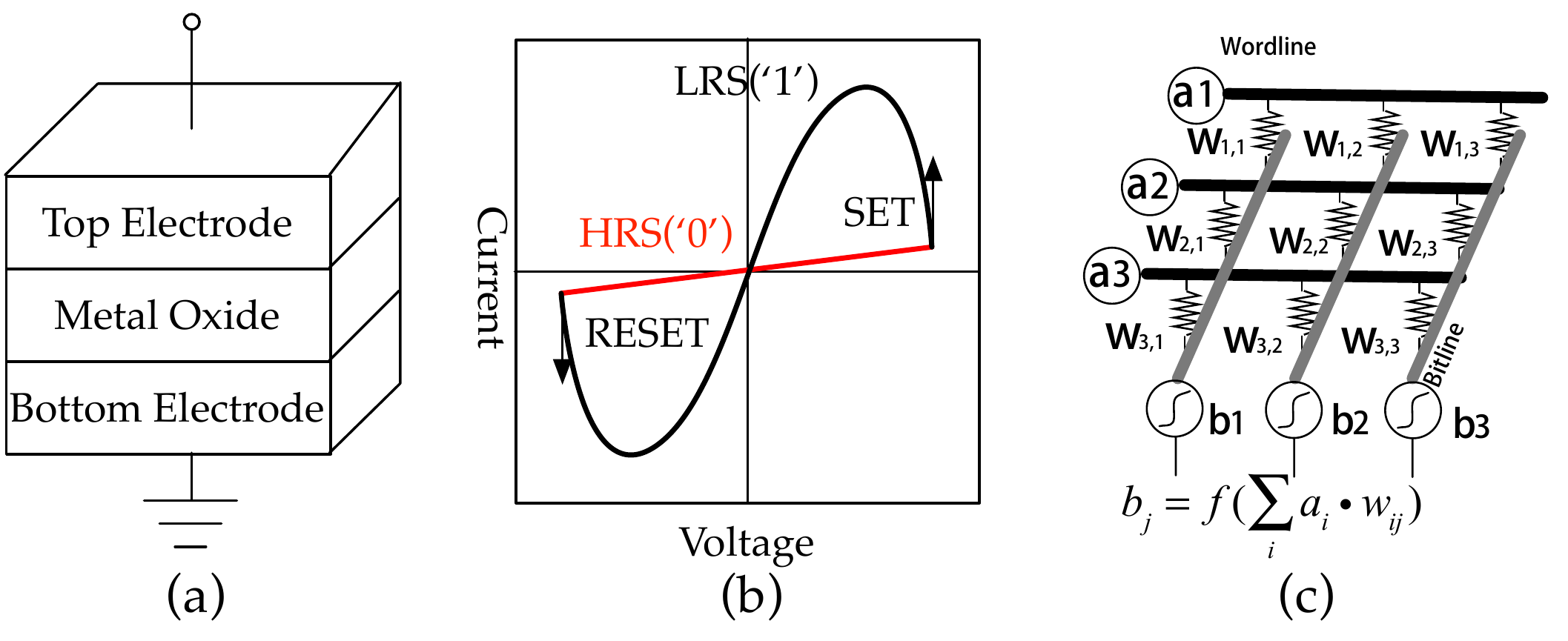}
\vspace{-5pt}
\caption{Basics of ReRAM }
\label{Fig_IVandCrossbar}
\vspace{-5pt}
\end{figure}

The resistive random access memory (ReRAM)~\cite{wong2012metal}
is an emerging non-volatile memory with appealing properties of 
high density, fast read access and low leakage power.
ReRAM has been considered as a promising candidate for future memory architecture. 
A primary application of ReRAM is to be used as an alternate for main memory~\cite{xu2015overcoming, liu2014130, fackenthal201419}. 
Figure~\ref{Fig_IVandCrossbar} (a) demonstrates the metal-insulator-metal (MIM) structure of an ReRAM cell. It has a top electrode, a bottom electrode and a metal-oxide layer sandwiched between electrodes. By applying an external voltage across it, an ReRAM cell can be switched between a high resistance state (HRS or OFF-state) and a low resistance state (LRS or On-state), which are used to represent the logical "0" and "1", respectively, as shown in Figure~\ref{Fig_IVandCrossbar} (b). The endurance of ReRAM could be up to $10^{12}$ \cite{lee2011fast, hsu2013self}, alleviating the lifetime concern faced by other non-volatile memory, such as PCM \cite{qureshi2009enhancing}.

The ReRAM features the capability to perform
{\em in-situ matrix-vector multiplication}~\cite{hu2016dot, hu2012hardware} as shown in Figure~\ref{Fig_IVandCrossbar} (c), which utilizes the property of bitline current summation in ReRAM crossbars to enable computing with high performance and low energy cost. 
While conventional CMOS based system showed success on neural network acceleration~\cite{chen2014dadiannao, mahajan2016tabla, albericio2016cnvlutin}, 
recent works~\cite{shafiee2016isaac, chi2016prime,pipelayer,liu2015reno} demonstrated that ReRAM-based architectures offer significant performance and energy benefits for the computation and memory intensive neural network computing.

\subsection{Graph Processing Accelerators}

Due to the wide applications of graph processing and its challenges, 
several hardware accelerators were recently proposed. 
Ahn {\em et al.}~\cite{ahn2015scalable} proposes \onurgraph, the first PIM-based graph processing architecture. 
It defines a generic communication interface to map graph processing to HMC. 
At any time, each core can {\tt Put} a remote memory access 
and get interrupted to receive and execute the {\tt Put} calls from other cores. 
Ozdal {\em et al.}~\cite{ozdal2016energy} introduces an accelerator for asynchronous graph processing,
which features efficient hardware scheduling and dependence tracking.
To use the system, programmers have to understand its architecture and modify existing code.
Graphicionado~\cite{hamgraphicionado} is a customized graph accelerator designed for high performance and energy efficiency based on off-chip DRAM and on-chip eDRAM. 
It modifies the graph data structure and data path to 
optimize graph access patterns and designs specialized memory subsystem for higher bandwidth. 
These accelerators all optimize the memory accesses, reducing the latency 
or better tolerating the random access latency.  
\cite{ahn2015pim} and \cite{nai2017graphpim} focused on the data coherence in memory which can be accessed by instructions from host CPU and the accelerator, and \cite{nai2017graphpim} also considered atomic operations.

\subsection{Graph Representation}
\label{graph_format}

As discussed in Section~\ref{_introduction}, 
supporting the matrix-vector multiplications on small 
data blocks could increase the ratio between computation and data movement and 
reduce the pressure on memory system. 
With its matrix-vector multiplication capability, ReRAM could naturally 
perform the low-cost parallel dense operations on the 
sparse sub-matrices (subgraphs), 
enjoying the benefits without increasing hardware and energy cost. 

The key insight of \scheme\ is to still store the majority of the graph data
in the compressed sparse matrix representation and 
process the subgraphs in uncompressed sparse matrix representation.
In the following, we review several commonly used sparse representations.

\begin{figure}[h]
\centering
\vspace{-9pt}
\includegraphics[width=0.95\columnwidth]{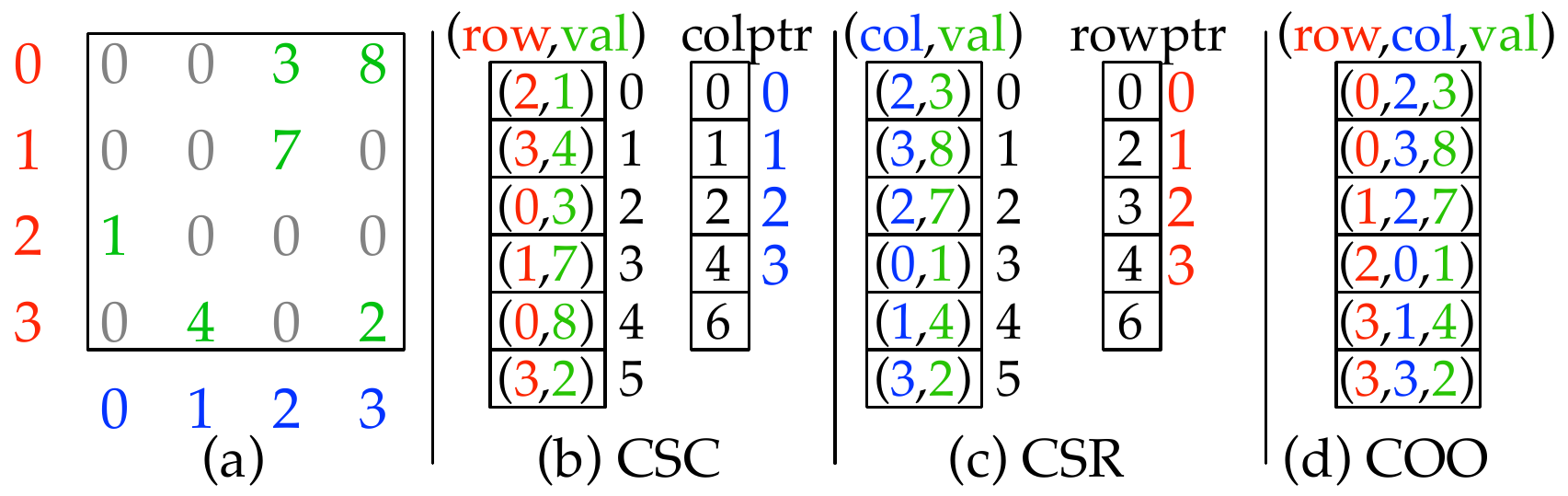}
\vspace{-9pt}
\caption{(a) Sparse Matrix and Its Compressed Representations in: (b) Compressed Sparse Column (CSC), (c) Compressed Sparse Row (CSR), (d) Coordinate List (COO)}
\label{sparse_form}
\vspace{-5pt}
\end{figure}

The three major compressed sparse representations are compressed sparse column (CSC), compressed sparse row (CSR) and coordinate list (COO). 
They are illustrated in Figure~\ref{sparse_form}. 
In the CSC representation, non-zeros are stored in column major order as (row index, value) pairs in a list, the number of entries in the list is the number of non-zeros. 
Another list of column starting pointers indicate the starting index of a row in the (row,val) list.
For example, in Figure~\ref{sparse_form} (a), {\bf 4} in the {\tt colptr} indicates that the {\bf 4}-th entry in (row,val) list, i.e., (0,8) is the starting of column 3.
The number of entries in {\tt colptr} is the number of columns + 1. Compressed sparse row (CSR) is similar to CSC, with row and column alternated. For coordinate list (COO), each entry is a tuple of (row index, column index, value) of the nonzeros.

\begin{figure}[htb]
\vspace{-5pt}
\centering
\includegraphics[width=\columnwidth]{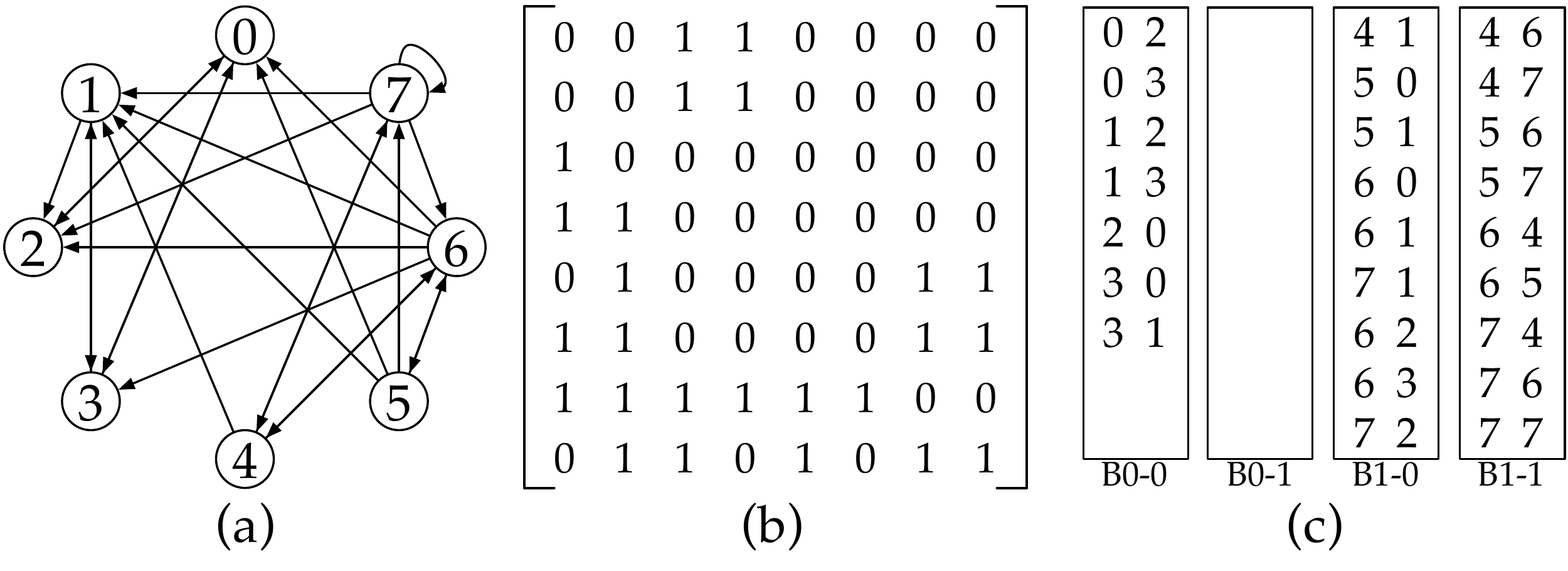}
\vspace{-20pt}
\caption{(a) A Directed Graph and Its Representations in (b) Adjacency Matrix and (c) Coordinate List}
\label{graph_rep}
\vspace{-5pt}
\end{figure}


In \scheme, we assume {\em coordinate list (COO)} representation to {\em store} a graph.
Given a graph in Figure~\ref{graph_rep}
\footnote{As shown in Section~\ref{graph_format}, each entry in a coordinate list should be a three-element tuple of \{Source ID, Destination ID, Edge Weight\}. To simplify the example,
we use an unweighted graph, so a two-element tuple of \{Source ID, Destination ID\} is 
sufficient to represent one edge.}
(a), its (sparse) adjacency matrix representation
and COO representation (partitioned into four 4 $\times$ 4 subgraphs) 
are shown in Figure~\ref{graph_rep}(b) and (c), respectively.
In this example, the coordinate list saves $61\%$ storing space,
compared with the adjacency matrix representation.
For real-world graphs with high sparsity, the saving is even high:
the coordinate list can only take $0.2\%$ of space for WikiVote~\cite{snapnets} compared to an adjacency matrix.




\section{GraphR Architecture}
\label{graphr_arch}

\subsection{Sparse Matrix Vector Multiplication (SpMV) and Graph Processing}
\label{arch_insights}

 \begin{figure}[h]
    \centering
	\vspace{-0.5cm}
\begin{lstlisting}[style=MyScala]
//Phase 1: compute edge values
for each edge E(V,U) from active vertex V do
	E.value = processEdge(E.weight,V.prop)
end for

//Phase 2: reduce and apply
for each edge E(U,V) to vertex V do
	V.prop = reduce(V.prop,E.value)
end for
\end{lstlisting}
	\vspace{-0.5cm}
	\caption{Vertex Programming Model}
	\vspace{-3mm}
	\label{fig:vertex}
\end{figure}

\begin{figure*}[t]
\centering
\includegraphics[width=1.9\columnwidth]{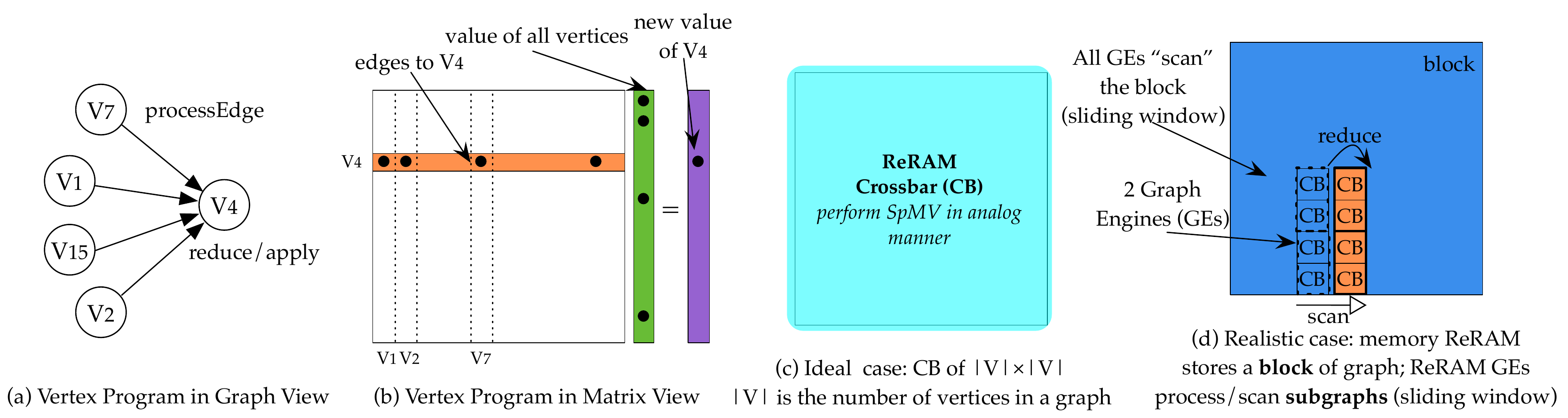}
\vspace{-10pt}
\caption{\scheme\ Key Insight: Supporting Graph Processing with ReRAM Crossbar}
\label{graphr_arch_insight}
\vspace{-8pt}
\end{figure*}

Figure~\ref{fig:vertex} shows the general vertex programming model for graph processing.
It follows the principle of ``Think Like The Vertex''~\cite{malewicz2010pregel}.  
In each iteration, the computation can be divided into two phases.
In the first phase, each edge is {\em processed} with \texttt{processEdge} function,
based on edge weight and active source vertex $V$'s property (\texttt{V.prop}), 
it computes a \texttt{value} for each edge. 
In the second phase, each vertex {\em reduces} values of all incoming edges (\texttt{E.value})
and generates a property value, which is {\em applied} to the vertex as its updated property.

Figure~\ref{graphr_arch_insight} (a) shows vertex program in graph view. 
$V_{15}$, $V_7$, $V_1$ and $V_2$ each executes a \texttt{processEdge} function, the results
are stored in corresponding edges to $V_4$.
After all edges are processed, $V_4$ executes \texttt{reduce} function to 
generate the new property and update itself. 
In graph processing, the operation in \texttt{processEdge} function is multiplication,
to generate the new property for each vertex, what essentially needed is 
the {\em Multiply-Accumulate (MAC)} operation.
Moreover, the vertex program of each vertex 
is equivalent to a sparse matrix vector multiplication (SpMV)
shown in Figure~\ref{graphr_arch_insight} (b).
Assume $A$ is the sparse adjacency matrix, and 
$x$ is a vector containing \texttt{V.prop} of all vertices,
the vertex program of {\em all vertices} can be computed 
{\em in parallel} in matrix view as $\mathbf{A^Tx}$.
As shown in Figure~\ref{Fig_IVandCrossbar} (c), ReRAM crossbar could perform 
matrix-vector multiplication efficiently.
Therefore, as long as a vertex program can be expressed in 
SpMV form, it can be accelerated by our approach. 
We will discuss in Section~\ref{graph_engine} on the
different patterns to express algorithms in SpMV. 

From Figure~\ref{graphr_arch_insight} (b), we see that the size of the matrix and vector 
are $|V|\times|V|$ and $|V|$, respectively, where $V$ is the number of vertices in the graph.  
In ideal case, if we are given a ReRAM crossbar (CB) of size $|V|\times|V|$, the vertex
program of each vertex can be executed in parallel and the new value (i.e., \texttt{V.prop}
in Figure~\ref{fig:vertex}) can be computed in one cycle.
More importantly, the memory to store $\mathbf{A^T}$ can directly
perform such in-memory computation without data movement. 
Unfortunately, this is unrealistic, the size of CB is quite limited 
(e.g., $4\times4$ or $8\times8$), 
to realize this idea, we need to {\em use Graph Engines (GEs) composed of 
small CBs to process subgraphs in certain manner}. 
This is the ultimate design goal of \scheme\ architecture. 
We face several key problems.
First, as discussed in Section~\ref{graph_format}, graph is stored in compressed
format, not in adjacency matrix, to perform in-memory computation, 
data needs to be converted to matrix format. 
Second, the real-world large graphs may not fit in memory. 
Third, the order of subgraph processing needs to be carefully determined,
because this affects the hardware cost to buffer the temporary results for reduction. 

To solve these problems, Figure~\ref{graphr_arch_insight} (d) 
shows the high level processing procedure. 
The \scheme\ architecture has both memory and compute module, each time, a {\em block}
of a large graph is loaded into \scheme's memory module (in blue) in 
compressed sparse representation.
If the memory module can fit the whole graph, then there is only one block, 
otherwise, a graph is partitioned into several blocks.
Inside \scheme, a number of GEs (in orange), each of which is composed of a number of CBs, 
``scan'' and process (similar to a {\em sliding window}) 
{\em subgraphs} in streaming-apply model (Section~\ref{streaming_apply}).
To enable in-memory computation, graph data are converted to matrix format by a 
controller. Such conversion is straightforward with preprocessed edge list (Section~\ref{preprocess}).
The intermediate results of subgraphs
are stored in buffers and simple 
computation logics are used to perform reduction. 

\scheme\ can be used in two settings:
{\em 1)} multi-node: one can connect different \scheme\ nodes and connect them 
together to process large graphs. In this case, each block is processed by a 
\scheme\ node.
Data movements happen between \scheme\ nodes.
{\em 2)} out-of-core: one can use one \scheme\ node as an in-memory graph processing
accelerator to avoid using multiple threads and moving data through memory hierarchy.
In this case, all blocks are processed consecutively by a single \scheme\ node.
Data movements happen between disk and \scheme\ node.
In this paper, we assume out-of-core setting, so we do not consider 
communication between nodes and its supports, 
we leave this as future work and extension.
Next, we present the overall \scheme\ architecture and its workflow.

\subsection{GraphR Architecture}
\label{arch_overall}

\begin{figure}[htb]
\vspace{-5pt}
\centering
\includegraphics[width=.92\columnwidth]{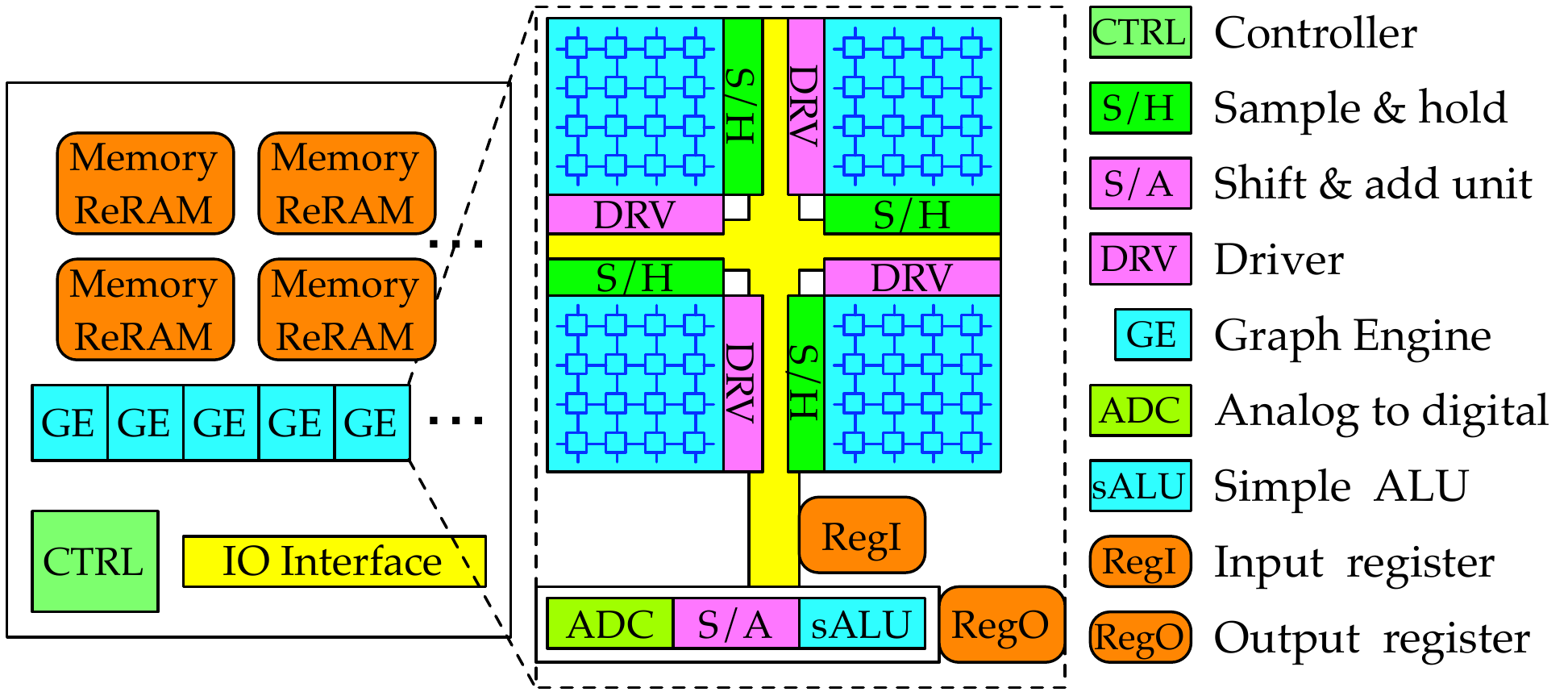}
\vspace{-5pt}
\caption{\scheme\ Architecture}
\label{arch_hier}
\vspace{-5pt}
\end{figure}



Figure~\ref{arch_hier} shows the architecture of one \scheme\ node.
\scheme\ is a ReRAM-based memory module that performs efficient 
near data parallel graph processing. 
It contains two key components: {\em memory ReRAM} and {\em graph engine (GE)}.
Memory ReRAM stores graph data in original compressed sparse representation.
GEs perform efficient
matrix-vector multiplications on matrix representation. 
A GE contains a number of ReRAM crossbars (CBs), {\em Drivers (DRVs)},
{\em Sample and Hold (S/H)} components placed in mesh, 
they are connected with {\em Analog to Digital Converter (ADC)}, {\em Shift and Add units (S/A)} and {\em simple algorithmic and logic units (sALU)}.
The input and output register ({\em RegI/RegO}) are used to cache data flow. 
We discuss the detail of several components as follows.

\textbf{Driver (DRV)}
It is used to {\em 1)} load new edge data to 
ReRAM crossbars for processing; and 
{\em 2)} input data into ReRAM crossbars for matrix-vector multiplication.



\textbf{Sample and Hold (S/H)}
It is used to sample analog values and hold them before converting to a digital form.


\textbf{Analog to Digital Converter (ADC)}
It converts analog values to digital format. Because ADCs have relatively higher area and power consumption, ADCs are not connected to every bitlines of ReRAM crossbars in a GE but shared between those bitlines. If the GE cycle is $64ns$, we can have one ADC working at $1.0GSps$ to convert all data from eight $8$-bitline crossbars within one GE. 

\textbf{sALU}
It is a simple customized algorithmic and logic unit. 
sALU performs operations that cannot be efficiently performed by ReRAM crossbars, such as comparison. 
The actual operation performed by sALU depends on 
algorithm and can be configured.
We will show more details when we discuss algorithms 
in Section~\ref{graph_engine}.

\textbf{Data Format}
It is not practical for ReRAM cells to support a high resolution.
Recent work~\cite{hu2016dot} reported 5-bit resolution on ReRAM programing. 
To alleviate the pressure of diver, we conservatively
assume the 4-bit ReRAM cell. 
To support higher computing resolution, e.g., 16 bit, 
the {\em Shift and Add (S/A)} unit is employed. A 16-bit fixed point number $M$ can be represented as $M=[M_3,M_2,M_1,M_0]$, where each segment $M_i$ is a 4-bit number. We can shift and add results from four 4-bit-resolution ReRAM crossbars, i.e. $D_3\ll12+D_2\ll8+D_1\ll4+D_0$ to get a 16-bit result.

When sALU and S/A are bypassed,
a graph engine could be simply considered as a memory ReRAM mat. 
A similar scheme of reusing ReRAM crossbars for computing and storing is employed in \cite{chi2016prime}.

 
The I/O interface is used to load
graph data and instructions into memory ReRAM and controller, respectively.
In \scheme, {\em controller} is the software/hardware interface that 
could execute simple instructions to:
{\em 1)} coordinate graph data movements between memory ReRAM and GEs 
based on streaming-apply execution model;
{\em 2)} convert edges in preprocessed coordinate list 
(assumed in this paper but can work with other representations as well) 
in memory ReRAM to sparse matrix format in GEs;
{\em 3)} perform convergence check. 

\begin{figure}[htb]
\centering
\vspace{-10pt}
\includegraphics[width=0.98\columnwidth]{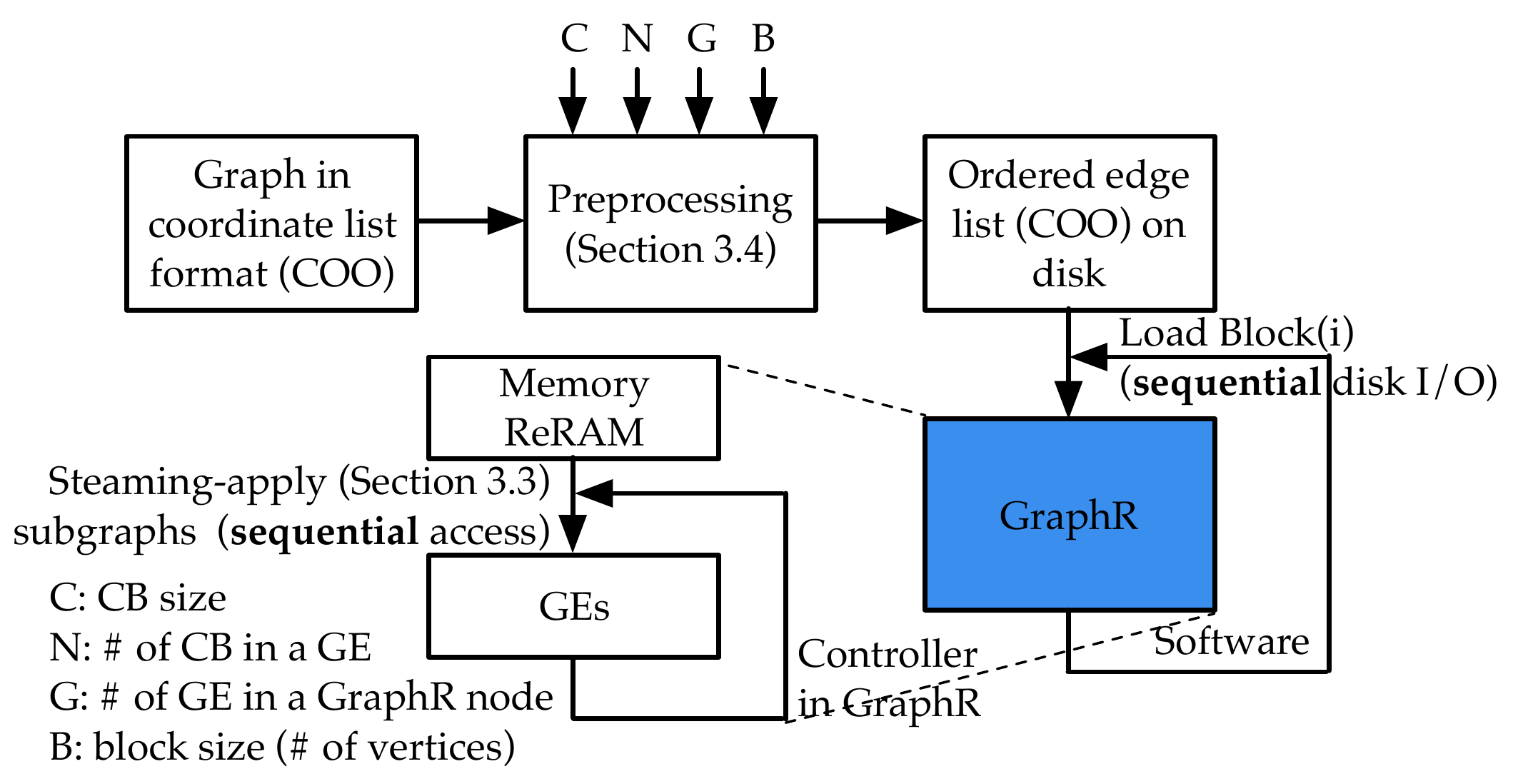}
\vspace{-5pt}
\caption{Workflow of \scheme\ in Out-Of-Core Setting}
\label{workflow}
\vspace{-5pt}
\end{figure}

 \begin{figure}[h]
    \centering
	\vspace{-0.0cm}
\begin{lstlisting}[style=MyScala]
while(true)
	load edges for next subgraph into GEs;
	process (processEdge) in GE;
	reduce by sALU;
	if(check_convergence())
         break;
}
	
\end{lstlisting}
	\vspace{-0.6cm}
	\caption{Controller Operations}
	\vspace{-3mm}
	\label{control_op}
\end{figure}

Figure~\ref{workflow} shows the workflow of \scheme\ in an out-of-core
setting. \scheme\ node can be used as a drop-in in-memory graph processing accelerator.
The loading of each block is performed in software by 
an out-of-core graph processing framework (e.g., GridGraph~\cite{zhu2015gridgraph}).
The integration is easy because it already contains the codes 
to load block from disk to DRAM, we just need to redirect data to \scheme\ node.
Since edge list is preprocessed in certain order, loading each block only involves
sequential disk I/O. 
Inside \scheme\ node, the data is initial loaded into memory ReRAM,  
the controller manages the data movements between GEs in streaming-apply manner. 
Edge list preprocessing for \scheme\ needs to be carefully designed and 
it is based on the architectural parameters.
Figure~\ref{control_op} shows the operations performed by 
controller.


\subsection{Streaming-Apply Execution Model}
\label{streaming_apply}

\begin{figure}[htb]
\centering
\vspace{-5pt}
\includegraphics[width=0.98\columnwidth]{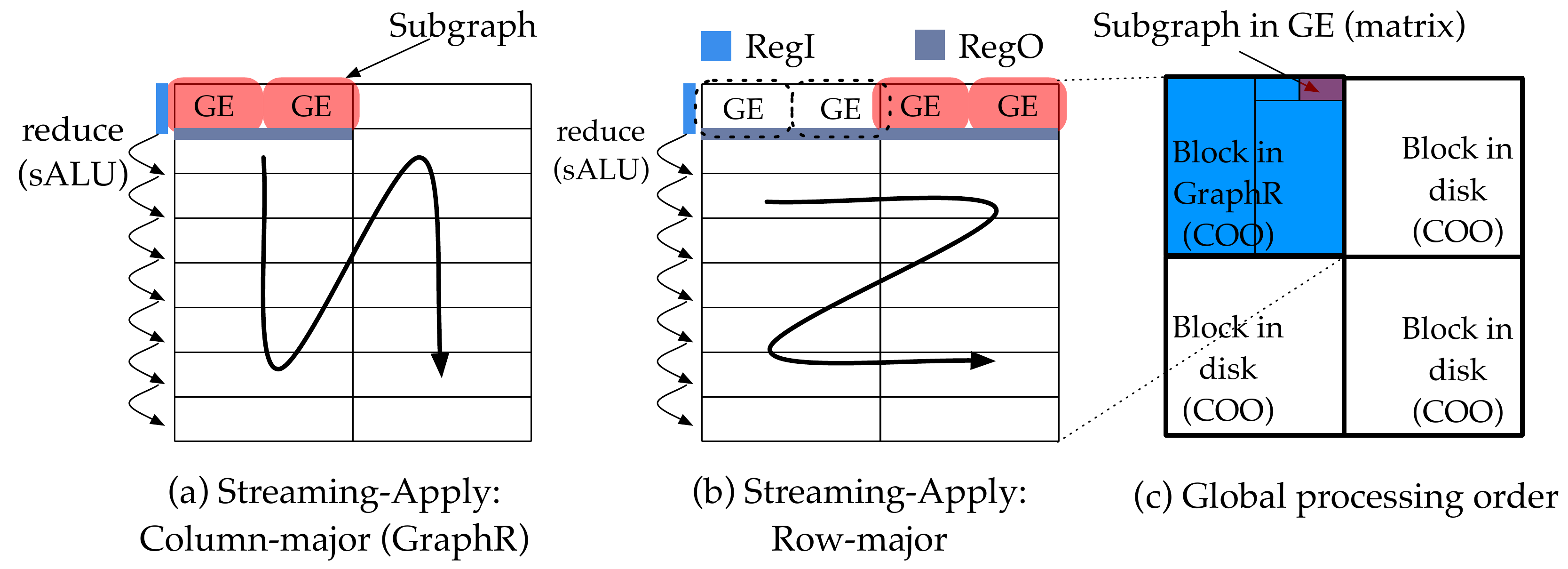}
\vspace{-5pt}
\caption{Streaming-Apply Execution Model}
\label{stream_apply}
\vspace{-5pt}
\end{figure}

In \scheme, all GEs process a subgraph, the order of processing is important, 
since it affects hardware resource requirement. 
We propose {\em streaming-apply} execution model shown in Figure~\ref{stream_apply}.
The key insight is that subgraphs are processed in a streaming manner and 
the reductions are performed on the fly by sALU. 
There are two variants of this model: column-major and row-major.
During execution, {\em RegI} and {\em RegO} are used to store source vertices 
and updated destination vertices. 
In column-major order in Figure~\ref{stream_apply} (a),
subgraphs with same destination vertices are processed together.  
The required {\em RegO} size is the same as the number of destination vertices
in a subgraph. 
In row-major order in Figure~\ref{stream_apply} (b),
subgraphs with same source vertices are process together.
The required {\em RegO} size is the total number of destination vertices of 
all subgraphs with the same source vertices. 
It is easy to see that row-major order requires larger {\em RegO}.
On the other side, row-major order incurs less read of {\em RegI}, --- only one read is 
needed for all subgraphs with same source vertices. 
In \scheme, we use column-major order since it requires less registers, 
and in ReRAM, the write cost is higher than read cost.

Figure~\ref{stream_apply} (c) shows the global processing order in a complete
out-of-core setting. We see that the whole graph is partitioned into 4 blocks, three of 
them are stored in disk in COO representation, one is stored in GraphR's memory ReRAM 
also in COO representation. 
Only the subgraph being processed by GEs is in sparse matrix format.
Due to the sparsity of graph data, if the subgraph is empty,
then GEs can move down to the next subgraph.
Therefore, the sparsity only incurs waste inside the subgraph
(e.g., when one GE has an empty matrix but others do not).

\subsection{Graph Preprocessing}
\label{preprocess}

To support streaming-apply execution model and conversion from 
coordination list representation to sparse matrix format, 
edge list needs to be preprocessed so that edges of consecutive 
subgraphs are stored together. It also ensures sequential 
disk and memory access on block/subgraph loading.  
In the following, we explain the preprocessing procedure in detail.

\begin{figure}[htb]
\centering
\vspace{-5pt}
\includegraphics[width=.85\columnwidth]{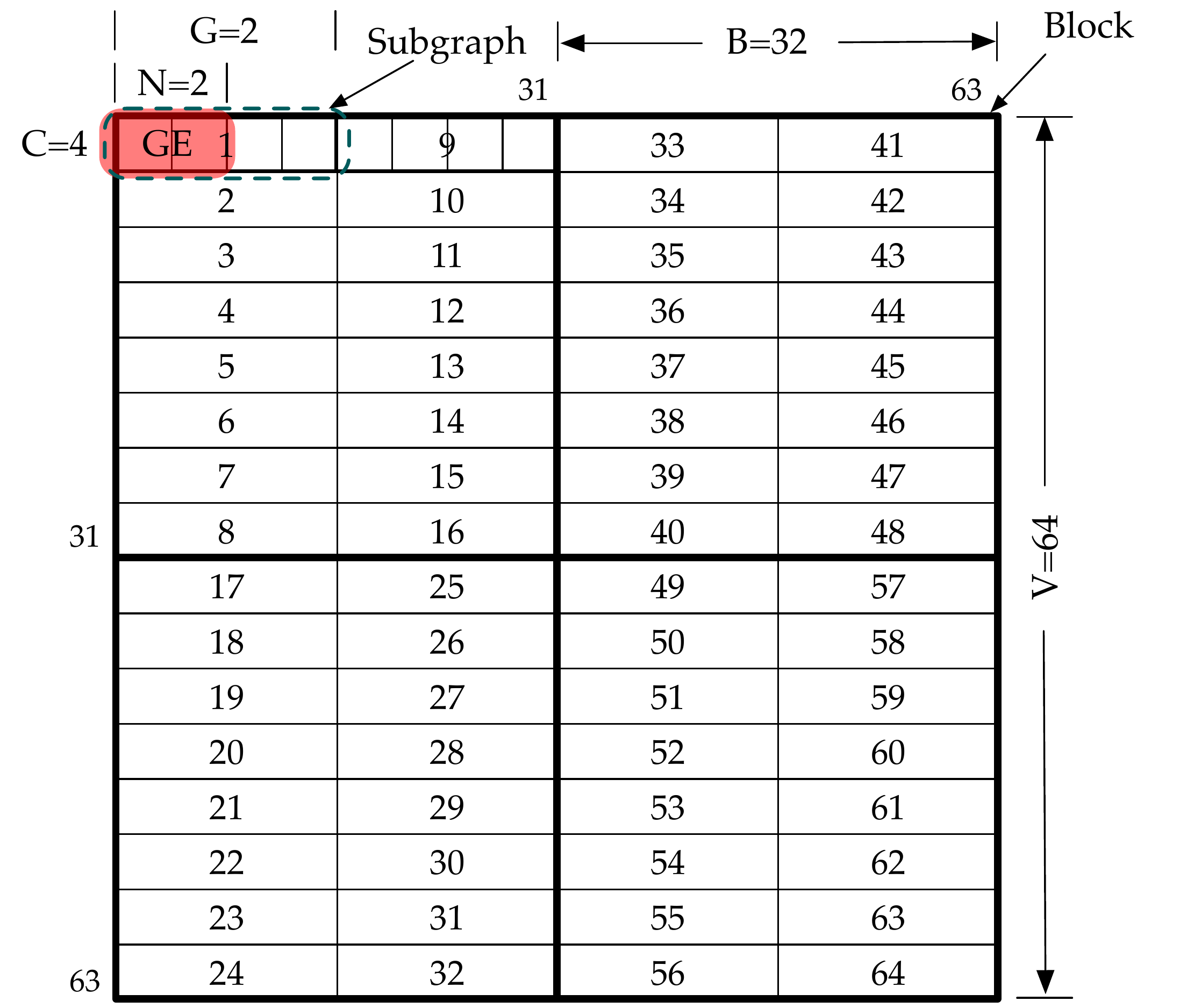}
\vspace{-10pt}
\caption{Preprocessing Edge List}
\label{preprocess}
\vspace{-5pt}
\end{figure}

Given some architectural parameters,
the preprocessing is implemented in software and performed only once.
Figure~\ref{preprocess} illustrates these parameters and 
the subgraph order we should produce. 
This order is determined by streaming-apply model.
$C$ is the size of CB;
$N$ is the number of CBs in a GE;
$G$ is the number of GEs in a \scheme\ node; and 
$B$ is the number of vertices in a block (i.e., block size).
Also, $V$ is the number of vertices in the graph. 
Among the parameters, $C$, $N$ and $G$ specify the architecture
setting of a \scheme\ node, $B$ is determined by the size of 
memory ReRAM in the node.
In the example, the graph has 64 vertices ($V=64$) 
and each block has 32 vertices ($B=32$), 
so the graph is partitioned into 4 blocks, each one will be loaded 
from disk to \scheme\ node.
Further, $C=4, N=2, G=2$, so the subgraph size is 
$C \times (C \times N \times G) = 4 \times (4 \times 2 \times 2) = 4 \times 16$.
Therefore, each block is partitioned into 16 subgraphs,
the number of each is the global subgraph order. 
Our goal is to produce an ordered edge list according this order
so that edges could be loaded sequentially.

Before presenting the algorithm, we assume that the original 
edge list is first ordered based on source vertex and then for 
all edges with the same source, they are ordered by destination. 
In another word, edges are stored in row-major order in matrix view. 
We also assume that in the ordered edge list, edges in a subgraph
is stored in column-major order.
Considering the problem in matrix view, each edge $(i,j)$, we first 
compute a global order ID ($I_{(i,j)}$), so we have a 3-tuple:
$(i,j,I_{(i,j)})$.
This global order ID takes all zeros into account, for example,
if there are $k$ zeros between two edges in the global order,
the different of global order ID of them is still $k$.
Then, all 3-tuples could be sorted by $I_{(i,j)}$. 
The ordered edge list is obtained if we output them according to 
this order. 
The space and time complexity of this procedure are
$O(V)$ and $O(VlogV)$, respectively, with common method. 
The key problem here is to compute $I_{(i,j)}$ for each $(i,j)$.
Without confusion, we simply denote $I_{(i,j)}$ as $I$.
We show that $I$ can be computed in a hierarchical manner. 

Let $B_I$ denote the global block order of the block that contains
$(i,j)$. We assume that blocks are also processed in column-major order:
$B_{(0,0)} \rightarrow B_{(1,0)} \rightarrow B_{(0,1)} \rightarrow B_{(1,1)}$.
The coordinates of a block B are:
\begin{equation} 
\small
\label{eq:block_co}
B_i=\floor*{\frac{i}{B}}, B_j=\floor*{\frac{j}{B}}
\end{equation} 
Based on column-major order, $B_I$ is:
\begin{equation} 
\small
\label{eq:block_id}
I_B = B_j + (V/B) \times B_j
\end{equation} 
We assume that $B$ can divide $V$, and similarly,
$C$ can divide $B$, $C \times N \times B$ can divide $B$. 
Otherwise, we can simply pad zeros to satisfy the conditions,
it will not affect the results since these zeros do not correspond to 
actual edges. The block corresponding to $B_I$ start with the following 
global order ID:
\begin{equation} 
\small
\label{eq:block_start}
start\_global\_ID(B_I) = B_I \times B^2 / (C^2 \times N \times G) + 1 
\end{equation}

Next, we compute $SI$, $(i,j)$'s global subgraph ID. 
The coordinates of the edge from the start of a block $B_I$ ($B_i,B_j$) are:
\begin{equation} 
\small
\label{eq:block_rel}
i' = i - B_i \times B, j' = j - B_j \times B
\end{equation}
The relative coordinates of the subgraph from the start of correspond 
block are:
\begin{equation} 
\small
\label{eq:block_rel_sg}
SI_{i'} = \floor*{\frac{i'}{C}}, SI_{j'} = \floor*{\frac{j'}{C \times N \times G}}
\end{equation} 
Then, we can compute $SI$:
\begin{equation}
\small
\label{eq:si}
	\begin{aligned}
SI & = (S_{i'} + S_{j'} \times B/C) + start\_global\_ID(B_I)   \\
& = (S_{i'} + S_{j'} \times B/C) + B_I \times B^2/(C^2 \times N \times G) + 1 
\end{aligned}
\end{equation}

Finally, we compute $SubI$, the relative order of $(i,j)$ from its
corresponding subgraph ($SI$). The coordinates are:
\begin{equation}
\small
\label{eq:subi_rel}
\begin{aligned}
SubI_i & = i - (B \times B_i) - (S_i \times C),\\
SubI_j & = j - (B \times B_j) - (S_j \times C)
\end{aligned}
\end{equation}
Since edges in a subgraph are assumed to be stored in 
column-major order, $SubI$ is:
\begin{equation} 
\small
\label{eq:subi}
SubI = SubI_i + (SubI_j - 1) \times C
\end{equation} 
With $SI$ and $SubI$ computed, we get $I$:
\begin{equation} 
\small
\label{eq:i}
I = (SI -1) \times (C^2 \times N \times G) + SubI
\end{equation} 

\subsection{Discussion}
\label{discuss}

\begin{figure}[htb]
\vspace{-0.5cm}
    \centering
\begin{lstlisting}[style=MyScala]
//Phase 1: compute edge values
for each edge E(V,U) from active vertex V do
	E.value = r * V.prop / V.outdegree
end for

//Phase 2: reduce and apply
for each edge E(U,V) to vertex V do
	V.prop = sum(E.value) + (1-r) / Num_Vertex
end for
\end{lstlisting}
	\vspace{-0.5cm}
	\caption{PageRank in Vertex Program}
	\vspace{-3mm}
	\label{fig:pr}
\end{figure}

\begin{figure}[htb]
    \centering
	\vspace{-0.2cm}
\begin{lstlisting}[style=MyScala]
//Phase 1: compute edge values
for each edge E(V,U) from active vertex V do
	E.value = E.weight + V.prop
	activate(U);
end forSomes

//Phase 2: reduce and apply
for each edge E(U,V) to vertex V do
	V.prop = min(V.prop,E.value)
end for
\end{lstlisting}
	\vspace{-0.5cm}
	\caption{SSSP in Vertex Program}
	\vspace{-5mm}
	\label{fig:sssp}
\end{figure}

\begin{table*}[t]
\vspace{-0pt}
\centering  
\small
\begin{tabular}[t]{||p{2cm}|p{1.7cm}|p{1.7cm}|p{1.7cm}|p{1.3cm}|p{3.2cm}|p{3.3cm}||}
\hline
 &CPU &GPU &Tesseract\cite{ahn2015scalable} &GAA\cite{ozdal2016energy}& Graphicionado\cite{hamgraphicionado} & {\bf \scheme}\\ \hline  
Process Edge & Instruction & Instruction & Instruction & Specialized AU &
Specialized unit &  ReRAM crossbar \\ \hline  
Reduce & Instruction & Instruction & Instruction and inter-cube communication & Specialized APU/SCU & Specialized unit & ReRAM crossbar or sALU \\
\hline 
Processing Model &Sync/Async           &Sync                     &Sync & Async  & Sync   &Sync \\ \hline 

Data Movement    &Disk to memory (out-of-core); \newline Memory hierarchy     
&Disk to memory; CPU/GPU memory; GPU mem. hierarchy    
&Between cubes (in-memory only)   
&Between memory and accelerator (in-memory only)  
&Between modules in memory pipeline; memory to SPM; SPM to/from processing units.
&Disk to memory (out-of-core) or 
between \scheme\ nodes (multi-node); \newline
Between memory ReRAM and GEs (inside \scheme)\\\hline
Memory Access &
\multicolumn{4}{|c|}
{ Random: vertex access and start of edge list of a vertex;}
 & Reduced random with SPM. & Sequential edge list. \\
& \multicolumn{4}{|c|} 
 { Sequential: edge list.} & Pipelined memory access & \\ \hline
Generality & \multicolumn{2}{|c|}{All algorithms} & 
\multicolumn{3}{|c|}{Vertex program} & Vertex program in SpMV \\ \hline 
\end{tabular}
\vspace{-0.3cm}
\caption{Comparison of Different Architectures for Graph Processing}
\vspace{-0.4cm}
\label{arch_compare}
\end{table*}

Table~\ref{arch_compare} compares different architectures
for graph processing. 
\scheme\ improves over previous architectures due to two unique features. 
First, the computation is performed in analog manner, others use
either instructions or specialized digital processing units. 
This provides the excellent energy efficiency. 
Second, all disk and memory accesses in \scheme\ are sequential,
this is due to preprocessing and less flexibility in scheduling 
vertices. It is a good trade-off because 
it is highly energy efficient to perform parallel operations 
in ReRAM CBs.
We believe that \scheme\ is the first architecture scheme using 
ReRAM CBs to perform graph processing, and the paper presents 
detailed and complete solution to integrate it as a drop-in accelerator
for out-of-core graph processing systems. 
The architecture, streaming-apply execution model and the preprocessing
algorithms are all novel contributions.
Also, \scheme\ is general because it could accelerate all 
vertex programs that can be performed in SpMV form.

\section{Mapping Algorithms in GE}
\label{graph_engine}

In this section, we discuss two patterns when mapping algorithms
to GEs: parallel MAC and parallel add-op. 
We use a typical example for each category
(i.e., PageRank and SSSP, respectively) to explain the insights. 
More examples (but not all) 
of supported algorithms in \scheme\ are listed in Table~\ref{app_operation_compare}. 
The first two are parallel MAC pattern and the 
last two are parallel add-op pattern.

\subsection{Parallel MAC}
\label{mapping_parallel}

In an algorithm, if \texttt{processEdge} function 
performs a multiplication, which can be performed in {\em each CB cell}, 
we call it {\em parallel MAC} pattern. 
The parallelization degree is roughly $(C \times C \times N \times G)$ (see parameter in Figure~\ref{workflow}).

PageRank~\cite{page1999pagerank} is an excellent example of this pattern.
Figure~\ref{fig:pr} shows its vertex program.
It does the following iterative computation:
$\overrightarrow{PR}_{t+1} = r\mathbf{M}\cdot\overrightarrow{PR}_{t} + (1-r)\overrightarrow{e}$.
$\overrightarrow{PR}_{t}$ is the PageRank at iteration $t$,
$\mathbf{M}$ is a probability transfer matrix, 
$r$ is the probability of random surfing and 
$\overrightarrow{e}$ is a vector of probabilities of staying in each page.




We consider a small subgraph that could be 
processed by a $5 \times 4$ CB (the additional row is to perform addition). 
It contains at most 16 edges represented by
the intersection of 4 rows and 4 columns in the sparse matrix (shown in Figure~\ref{twoexamples} (a)).
Thus, the block is related to 8 vertices (i.e., $i0 \sim i3$, $j0 \sim j3$).
The following are the parameters for the 4 $\times$ 4 block PageRank shown in Figure~\ref{twoexamples} (b1):
$\mathbf{M} = 
[
   0 ,   1/2 , 1 , 0 ;
   1/3 ,   0 , 0 , 1/2 ; 
   1/3 ,   0 , 0 , 1/2 ; 
   1/3 , 1/2 , 0 , 0
  ]$
,$\overrightarrow{e} =
  [
    1/4 , 1/4 , 1/4 , 1/4
  ]^\text{T}$,
$r=4/5$.

\begin{figure}[htb]
\vspace{-8pt}
\centering
\includegraphics[width=0.85\columnwidth]{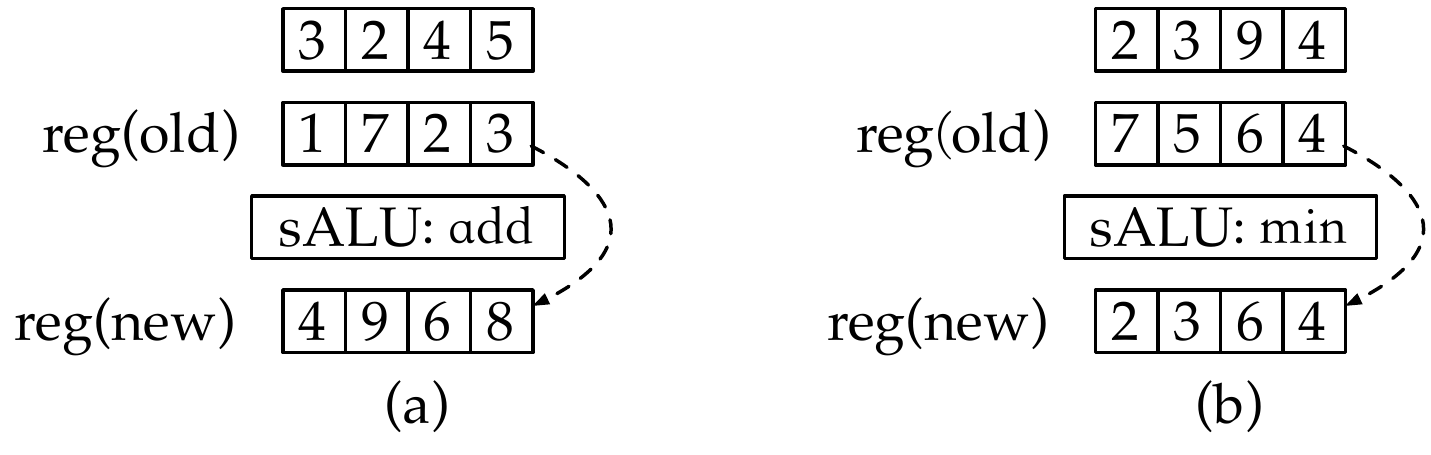}
\vspace{-13pt}
\caption{sALU Is Configured to Perform (a) {\tt add} in PageRank and (b) {\tt min} in SSSP}
\label{reg_old_new}
\vspace{-10pt}
\end{figure}

\begin{figure*}[t]
\centering
\includegraphics[width=1.9\columnwidth]{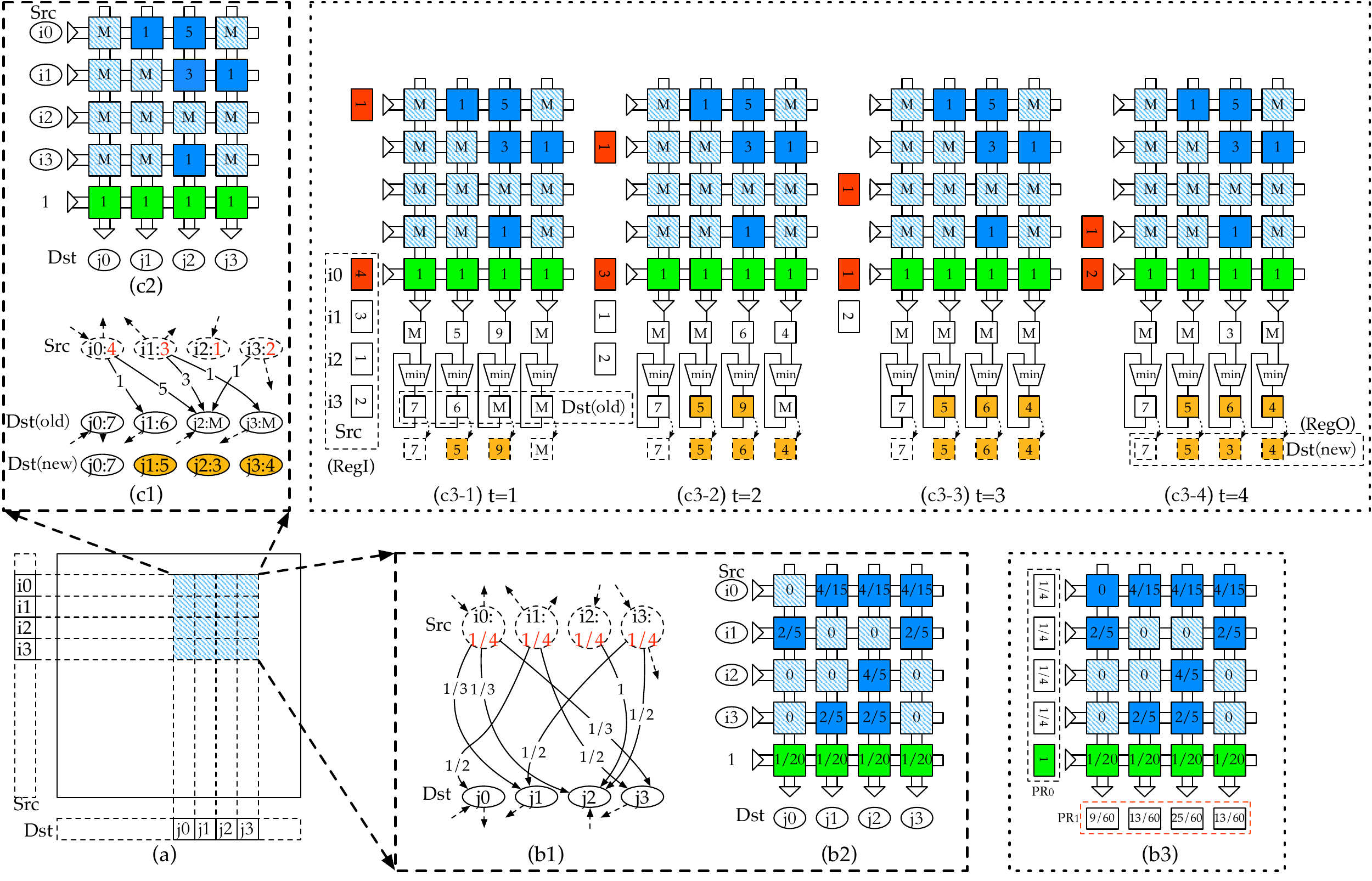}
\vspace{-10pt}
\caption{Graphs to Illustrate (a) PageRank and (b) SSSP}
\label{twoexamples}
\vspace{-10pt}
\end{figure*}

We define $\mathbf{M_0}=r\mathbf{M}$ and $\overrightarrow{e_0}=(1-r)\overrightarrow{e}$, so that $\overrightarrow{PR}_{t+1} = \mathbf{M_0}\overrightarrow{PR}_{t} + \overrightarrow{e_0}$.
In CB in Figure~\ref{twoexamples} (b2, b3), 
the values are already scaled with $r$. 
Figure~\ref{twoexamples} (b3) shows the mapping of PageRank algorithm 
to a CB. The additional row is used to
implement the addition of $\overrightarrow{e_0}$.
The sALU is configured to perform {\tt add} operation 
in the \texttt{reduce} function to add PageRank values (Figure~\ref{reg_old_new} (a)).
To check convergence, 
the new PageRank value is compared with it in the previous iteration, 
the algorithm is converged if the difference is less than a threshold. 

\subsection{Parallel Add-Op}
\label{mapping_serial}

In an algorithm, if \texttt{processEdge} function 
performs an addition, which can be performed in for 
{\em each CB row} at a time, 
we call it {\em parallel add-op} pattern.
The {\em op} specifies the operation in \texttt{reduce} that 
is implemented in sALU. 
The parallelization degree is roughly $(C \times N \times G)$.
Single Source Shortest Path (SSSP)~\cite{cormen2009introduction} 
is a typical example of this pattern. 
Figure~\ref{fig:sssp} shows the vertex program. 
We see that \texttt{processEdge} performs an addition and 
\texttt{reduce} performs {\em min} operation.
Therefore, sALU is configured to perform {\tt min} operation shown in Figure~\ref{reg_old_new} (b).


In SSSP, 
each vertex $v$ is given a distance label
$dist(v)$
that maintains the length of the shortest known path to vertex $v$ from the source. 
The distance label is initialized to 0 at the source 
and $\infty$ at all other nodes. 
Then the SSSP algorithm iteratively applies the 
{\em relaxation operator}, which is defined as follows: 
if $(u,v)$ 
is an edge and $dist(u) + w(u,v) < dist(v)$,
the value of $dist(v)$ is updated to 
$dist(u) + w(u,v)$.
An active vertex is relaxed by applying this operator to all the edges connected to it. 
Each relaxation may lower the distance label of some vertex, 
and when no further lowering can be performed anywhere in the graph, 
the resulting vertex distance labels indicate the shortest distances 
from the source to the vertex,
regardless of the order in which the relaxations were performed. 
Breadth-first numbering of a graph is a special case of SSSP 
where all edge labels are 1.




We explain the mapping of SSSP algorithm to a CB using a small 8-vertex
subgraph corresponding to a 4 $\times$ 4 block in sparse matrix, as shown in Figure~\ref{twoexamples} (c1). 
It can be represented by adjacency matrix: $\mathbf{W}=[
   M , 1 , 5 , M ;
   M , M , 3 , 1 ; $ 
   $
   M , M , M , M ; 
   M , M , 1 , M
  ]$
where $M$ indicates no edge connected two vertices and $M$ is set to a 
reserved maximum value for a memory cell in a CB.
The values are stored in the CB shown in  
Figure~\ref{twoexamples} (c2).

Given a vertex $u$ and $dist(u)$, the row in the adjacency matrix $\mathbf{W}$ 
for $u$ indicates $w(u,v)$.
We could {\em perform the relaxation operator (i.e., addition)
for $u$ in parallel}.
Here, SpMV is only used to select a row in CB by 
multiplying with an one-hot vector. 
 
The relaxation operator of $u$ involves {\em reading}:
{\em 1)} $dist(u)$: it is computed iteratively from the source, for 
the source vertex, the initial value is {\em zero};
{\em 2)} The {\em vector} of the $dist(v)$ before the relaxation operator: 
it is a vector indicating the distance between source and all other vertices and
is also computed iteratively from the source.
In our example, for the source vertices $(i0, i1, i2, i3)$, 
the initial value is $[4,3,1,2]$;
{\em 3)} The {\em vector} of the $w(u,v)$: it is 
the distance from $u$ to the destination vertices in the subgraph, and can be obtained 
by reading a row in adjacency matrix $\mathbf{W}$. 

Figure~\ref{twoexamples} (c3) shows the process to perform SSSP in a 5 $\times$ 4 CB.
The last row (\textcolor{green}{green} squares) is set to a fixed value 1, 
which is used to add $dist(u)$ (the input to the last wordline)
to each $w(u,v)$ in the relaxation operator. 
The initial value for $dist(u)$ for the destination vertices $(j0, j1, j2, j3)$ are $[7,6,M,M]$.
The vector of $w(u,v)$ is obtained by activating the wordline associated to vertex $v$.
In the time slot ($t=1$), wordline 0 (for source vertex $i0$) is activated 
(the \textcolor{red}{red} square with input value 1) 
and a value 4 (this is the current value in $dist(v)$ for source $i0$)
is input to the last wordline (the green box line).
The vector of the $dist(v)$ for the destination vertices 
($(j0,j1,j2,j3)$) is set as the value of the output
at each bitline, which is $[7,6,M,M]$. 
With this mapping, the current summation in bitline in Figure~\ref{twoexamples} (c3-1)
is $[1\times M+4\times1,1\times1+4\times1,1\times5+4\times1,1\times M+4\times1]
=[M,5,9,M]$.
It is the $dist(u) + w(u,v)$ computed in parallel, where $u$ is the source vertex. 
Then the distance of source to each vertex $v$ is compared with the 
initial $dist(v)$ ($[7,6,M,M]$) by an array of comparators, 
and in the final output of bitline, we get $[7,5,9,M]$, which is the 
updated $dist(v)$ vector after time slot $t=1$.
The parallel comparisons are performed by vertex-related operations in. 
Different algorithms may require different functions on vertices.

\begin{table*}[htbp]
\centering 
\small
\begin{tabular}{||l|l|l|l|l||}
\hline
Applications & Vertex Property      & processEdge()                             & reduce() & Ative Vertex List \\ \hline \hline 
SpMV         & Multiplication Value & E.value = V.prop / V.outdegree * E.weight & V.prop = sum(E.value)         & Not Required      \\ \hline
PageRank     & Page Rank Value      & E.value = r * V.prop / V.outdegree        & V.prop = sum(E.value) + (1-r) / Num\_Vertex & Not Required      \\ \hline
BFS          & Level                & E.value = 1 + V.prop                      & V.prop = min(V.prop,E.value)  & Required      \\ \hline
SSSP         & Path Length          & E.value = E.weight + V.prop               & V.prop = min(V.prop,E.value)  & Required      \\ \hline
\end{tabular}
\vspace{-0.2cm}
\caption{Property and Operations of Applications in \scheme}
\vspace{-0.45cm}
\label{app_operation_compare}
\end{table*}

After time slot $t=1$, we move to the next vertex in Figure~\ref{twoexamples} (c3-2), 
where we {\em i)} activate the second wordline;
{\em ii)} change the input to the last wordline to 3 (the distance label for source vertex i1); 
and {\em iii)} set the intermediate $dist(v)$ to be the final output of bitline in time slot
$t=1$, which is $[7,5,9,M]$. 
Similar as Figure~\ref{twoexamples} (c3-1), the 
the current summation in bitline is $[M,M,6,4]$ and it is compared 
with $[7,5,9,M]$, yielding the final output of bitline for 
time slot $t=2$ as $[7,5,6,4]$.
We indicate the updated distance label using \textcolor{orange}{orange} squares. 
The operations in time slot $t=3$ and $t=4$ can be performed in the similar manner.

Initially, before processing the block, 
the active indicator for each destination vertex is set to be FALSE. After the serial processing in CB, 
the active indicators for all vertices which have been updated  
(marked in \textcolor{orange}{orange} in Figure~\ref{twoexamples} (c3)) are set to be TRUE.
This indicates that they are active for next iteration. 
In our example, $j1, j2, j3$ are marked active.
Referring to Figure~\ref{twoexamples}, this means that the distance labels for 
these vertices have been updated. Because we are processing the block in CB, 
the active indicator for destination vertex may be accessed for multiple times, but if it is set be TRUE at least one time, this vertex is active in next iteration. 
Globally, after all active source vertices and corresponding edges are processed in an iteration, 
source vertex properties (values and active indicators) that hold the old values
are updated by the properties of the same vertices in destination.
The new source vertex properties are used in the next iteration.
We can check if there are still active vertices to determine the convergence.

\section{Evaluation}
\label{_evaluation}
\subsection{Graph Datasets and Applications}

\begin{table}[htb]
\vspace{-5pt}
\centering 
\small
\begin{tabular}{||c|c|c||}
\hline
Dataset &\# Vertices&\#Edges \\ \hline  
WikiVote(WV)~\cite{snapnets} &7.0K &103K 
 \\ \hline     
Slashdot(SD)~\cite{snapnets} &82K &948K 
 \\ \hline   
Amazon(AZ)~\cite{snapnets} &262K &1.2M 
 \\ \hline
WebGoogle(WG)~\cite{snapnets} &0.88M &5.1M 
 \\ \hline
LiveJournal(LJ)~\cite{snapnets} &4.8M &69M 
 \\ \hline
Orkut(OK)~\cite{nr} &3.0M &106M 
 \\ \hline
Netflix(NF)~\cite{bennett2007netflix} &480K users, 17.8K movies &99M 
 \\ \hline
\end{tabular}
\vspace{-8pt}
\caption{Graph Datasets}
\vspace{-12pt}
\label{graph_dataset}
\end{table}

Table~\ref{graph_dataset} shows the datasets used in our evaluation. 
We use seven real-world graphs. 
For WikiVote(WV), Slashdot(SD), Amazon(AZ), WebGoogle(WG), LiveJournal(LJ), Orkut(OK) and Netflix(NF). We run pagerank(PR), breadth first search(BFS), single source shortest path (SSSP) and sparse matrix-vector multiplication (SpMV) on the first six datasets. On Netflix(NF), we run collaborative filtering(CF), and the feature length used is 32.

\subsection{Experiment Setup}

In our experiments, we compare \scheme\ with a CPU
baseline platform, a GPU platform and PIM-based architecture~\cite{ahn2015scalable}.
PR, BFS, SSSP and SpMV running on the CPU platform are based on the software framework GridGraph~\cite{zhu2015gridgraph}, while collaborative filtering is based on GraphChi~\cite{kyrola2012graphchi}. PR, BFS, SSSP and SpMV running on GPU platform are based on Gunrock~\cite{wang2016gunrock}, while CuMF\_SGD~\cite{xie2016cumf_sgd} is the GPU framework for CF.
We evaluate PIM-based architecture on zSim~\cite{Sanchez-zsim}, a scalable x86-64 multicore simulator. We modified zSim with HMC memory and interconnection model, heterogeneous compute units, on-chip network and other hardware features. The results are validated results with NDP~\cite{gao2015practical}, which also extends zSim for HMC simulation.
In all experiments, graph data could fit in memory.
We also exclude the disk I/O time from the execution time of the CPU/GPU-based platform.

\begin{table}[htb]
\small
\vspace{-4pt}
\centering 
\begin{tabular}{||l|l||}
\hline
CPU &Intel Xeon E5-2630 V3,\\
 & 8 cores,  2.40 GHz\\
 &$8\times (32+32)$KB L1 Cache \\ 
 &$8\times 256$KB L2 Cache \\ 
 &20 MB L3 Cache \\ \hline  
Memory &128 GB \\ \hline     
Storage &1 TB \\ \hline   
\multicolumn{2}{||l||}
{
2 CPUs, a total number of 32 threads.
}\\
\hline
\end{tabular}
\vspace{-8pt}
\caption{Specifications of the CPU Platform}
\vspace{-10pt}
\label{cpuconfigure}
\end{table}

\begin{table}[htb]
\small
\vspace{-3pt}
\centering 
\begin{tabular}{||l|l||}
\hline
Graphic Card &NVIDIA Tesla K40c\\ \hline  
Architecture &Kepler\\ \hline 
\# CUDA Cores &2880\\ \hline 
Base Clock &745 MHz\\ \hline 
Graphic Memory &12 GB GDDR5 \\ \hline   
Memory Bandwidth &288 GB/s\\ \hline 
CUDA Version &7.5\\ \hline 
\end{tabular}
\vspace{-10pt}
\caption{Specifications of the GPU Platform}
\vspace{-10pt}
\label{gpuconfigure}
\end{table}

Specifications of the CPU and GPU platforms are shown in Table~\ref{cpuconfigure} and Table \ref{gpuconfigure}. The CPU energy consumption is estimated by Intel Product Specifications~\cite{intelspecifications} while NVIDIA System Management Interface ({\tt nvidia-smi}) is used to estimate energy consumption by GPU. The execution times for CPU/GPU platform are measured in the computing frameworks.

To evaluate \scheme, for the ReRAM part, we use NVSim \cite{dong2012nvsim} to estimate time and energy consumption.
The HRS/LRS resistance are 25M/50K $\Omega$, read voltage (Vr) and write voltage (Vw) are 0.7V and 2V respectively, and current of LRS/HRS are 40 uA and 2 uA respectively.
The read/write latency and read/write energy cost used are 29.31ns / 50.88ns, 1.08 pJ / 3.91nJ respectively from data reported in~\cite{niu2013design}.
The programming of a bipolar ReRAM cell is to change (from High to Low) or inverse. For multi-level cell, the programming is to change the resistance to a middle state between High and Low, and the middle state is determined by the programming
voltage pulse length. Actually, the difference between High and Low is the {\em worse case}. 
Note that ~\cite{hu2016dot,alibart2012high} describe a ReRAM programming prototype, 
which includes: 1) writing circuitry; 2) ReRAM cell/array; and 3) conversion circuitry. 
They demonstrated the {\em possibility} of 1\% accuracy for multi-level cell. 
However, in a real production system, only ``writing circuitr'' and ``ReRAM cell/array'' are needed, there is no need to consider the conversion, as we just need to ``acquiesce'' a writing precision. Therefore, this energy cost estimation for 4-bit cell programming is reasonable and more 
conservative than two recent ReRAM-based accelerators~\cite{shafiee2016isaac,chi2016prime}.

For on-chip registers, we use CACTI 6.5~\cite{CACTI} at 32nm to model. For Analog/Digital converters, we use data from~\cite{adcperf}. The system performance is modeled by code instrumentation. The ReRAM crossbar size $S$, number of ReRAM crossbars per graph engine $C$ and number of graph engines is 8, 32, 64 respectively.

\subsection{Performance Results}
\label{eval_perf}



\begin{figure}[htb]
\vspace{-10pt}
\centering
\includegraphics[width=0.99\columnwidth]{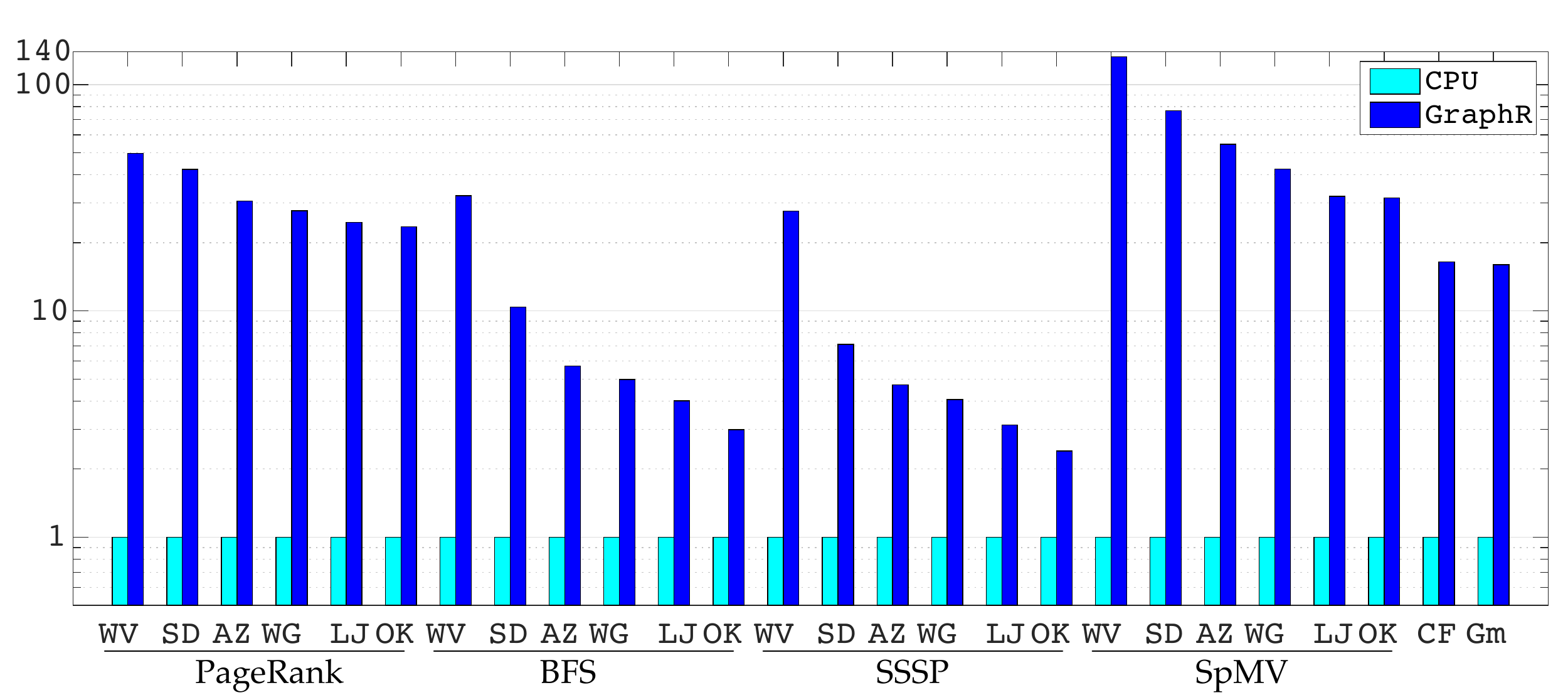}
\vspace{-10pt}
\caption{\scheme\ Speedup Compared to CPU}
\label{speedup}
\vspace{-10pt}
\end{figure}

Figure~\ref{speedup} compares the performance of 
\scheme\ and CPU platform. 
The CPU implementation is used as the baseline and execution times of
applications of \scheme\ are normalized to it.
Compared to CPU platform, the geometric mean of speedups  
with \scheme\ architecture on all 25 executions is
16.01$\times$.
Among all applications on the datasets, the highest speedup achieved by \scheme\ is 132.67$\times$, and it happens on SpMV on WikiVote dataset. The lowest speedup \scheme\ achieved is 2.40$\times$, on SSSP using OK dataset.
PageRank and SpMV are parallel MAC pattern and have higher 
speedup compared to CPU-based platform. 
For BFS and SSSP which are parallel add-op pattern, 
\scheme\ achieves lower speedups only due to parallel addition. 


\subsection{Energy Results}
\label{eval_energy}


\begin{figure}[htb]
\vspace{-12pt}
\centering
\includegraphics[width=0.99\columnwidth]{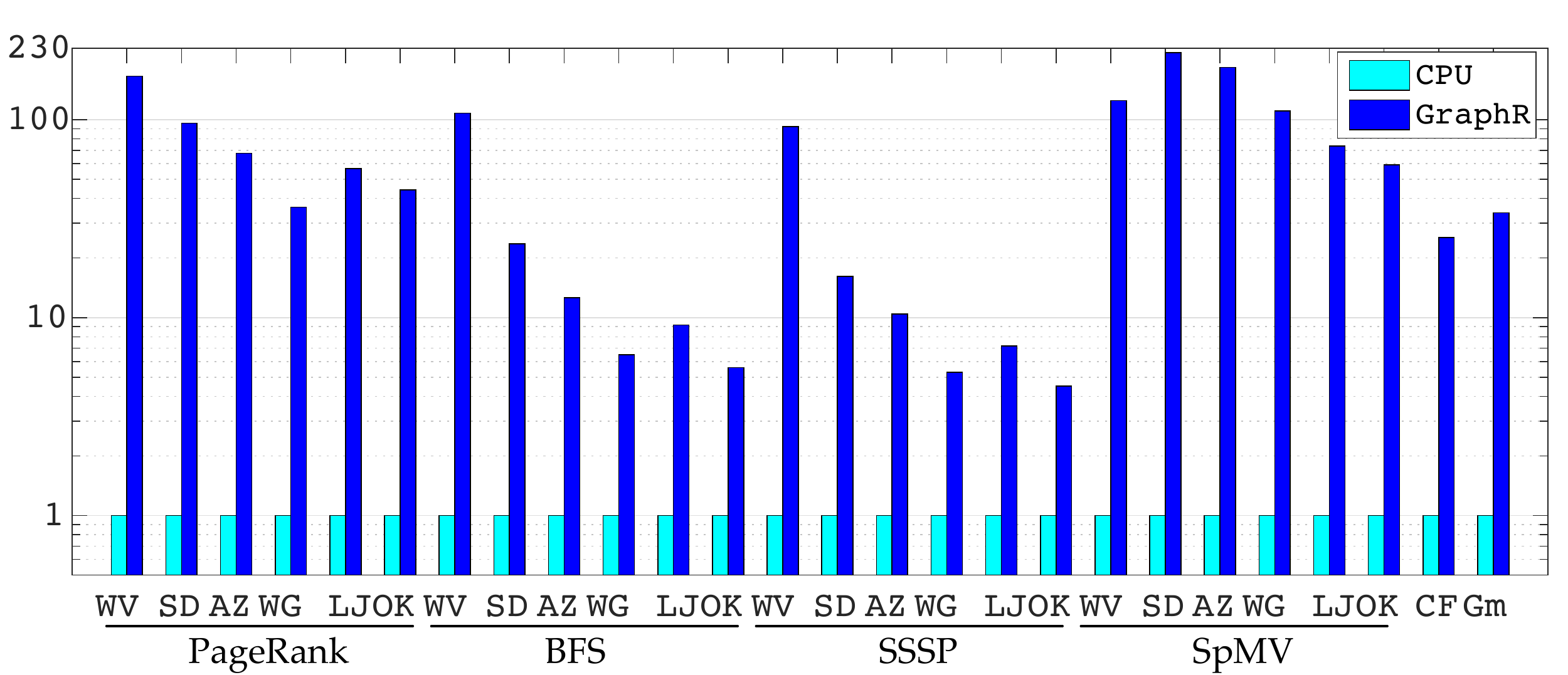}
\vspace{-10pt}
\caption{\scheme\ Energy Saving Normalized to CPU Platform}
\label{f_energy}
\vspace{-10pt}
\end{figure}

Figure \ref{f_energy} shows the energy savings in \scheme\
architecture over CPU platform.
The geometric mean of energy savings of all applications compared to CPU
is 33.82$\times$.
The highest energy achieved by \scheme\ is 217.88$\times$, which is on sparse matrix-vector multiplication on SD dataset.
The lowest energy saving achieved by \scheme\ happens on SSSP on OK dataset, which is 4.50$\times$. \scheme\ gets the high energy efficiency
from the non-volatile property of ReRAM 
and the in-situ computation capability.

\subsection{Comparison to GPU Platform}

GPUs take advantage of a large amount of threads
(CUDA cores) for high parallelism. The GPU used in the comparison has 2880 CUDA cores, while in \scheme\ we have a comparable number ($2048=32\times 64$) of crossbars. 
To compare with GPU, we run PageRank and SSSP on LiveJournal dataset, 
and collaborative filtering(CF) on Netflix dataset.

\begin{figure}[htb]
\vspace{-0pt}
\centering
\includegraphics[width=0.98\columnwidth]{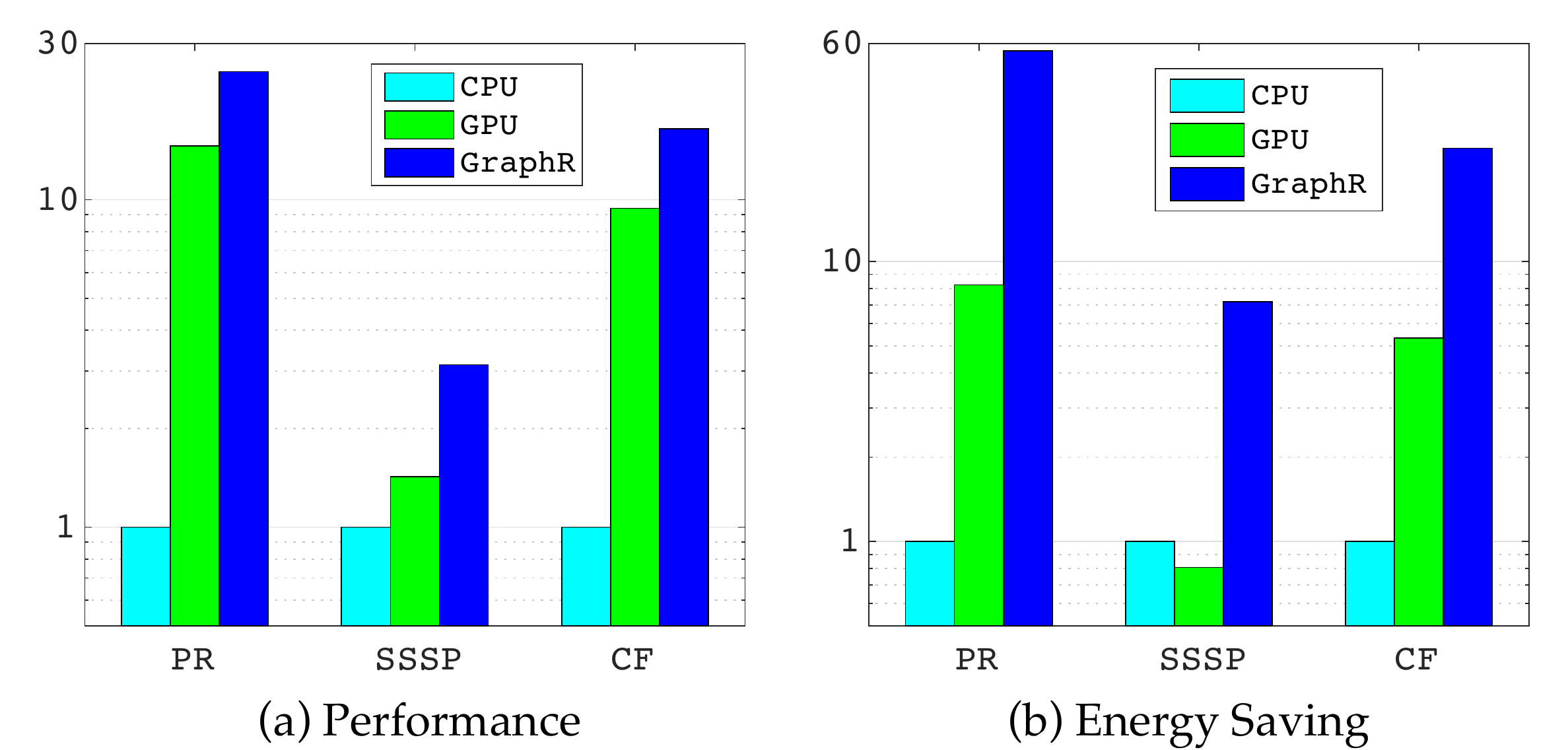}
\vspace{-10pt}
\caption{\scheme\ (a) Performance and (b) Energy Saving Compared to GPU Platform}
\label{togpu}
\vspace{-15pt}
\end{figure}

The performance and energy saving normalized to CPU are shown in Figure~\ref{togpu} (a) and (b).
Overall, the performance of \scheme\ is higher than GPU with {\em considering
the data transfer time between CPU memory and GPU memory}, --- an overhead \scheme\ does not incur. \scheme\ has 
performance gains ranging from 1.69$\times$ to 2.19$\times$ compared to GPU.
More importantly, \scheme\ consumes 4.77$\times$ to 8.91$\times$ less energy than GPU.
Figure~\ref{togpu} (a) shows that, \scheme\ achieves higher speedups on PageRank and CF, where MACs dominate the computation and are fully supported by \scheme\ and GPU. 
For SSSP, the vertex-related traversing in \scheme\ requires accessing to main memory and storage. 
In GPU, a cache based memory hierarchy better supports the random accessing. So the speedup of \scheme\ is lower. 
The reason why \scheme\ still has gain in SSSP is due to 
 sequential access pattern.
For energy saving, \scheme\ is better than GPU. Besides energy saving by in-situ computation in GEs and the in-memory processing system design, from Fig 17 in \cite{hamgraphicionado} we see that in conventional CMOS system, static energy consumption by eDRAM (memory) incurs the majority of energy consumption. As the technology node scales down, leakage power dominates in CMOS system. In contrast, ReRAM has almost 0 static energy leakage, so \scheme\ has higher energy saving.

\subsection{Comparison to PIM Platform}
\begin{figure}[htb]
\vspace{-12pt}
\centering
\includegraphics[width=0.98\columnwidth]{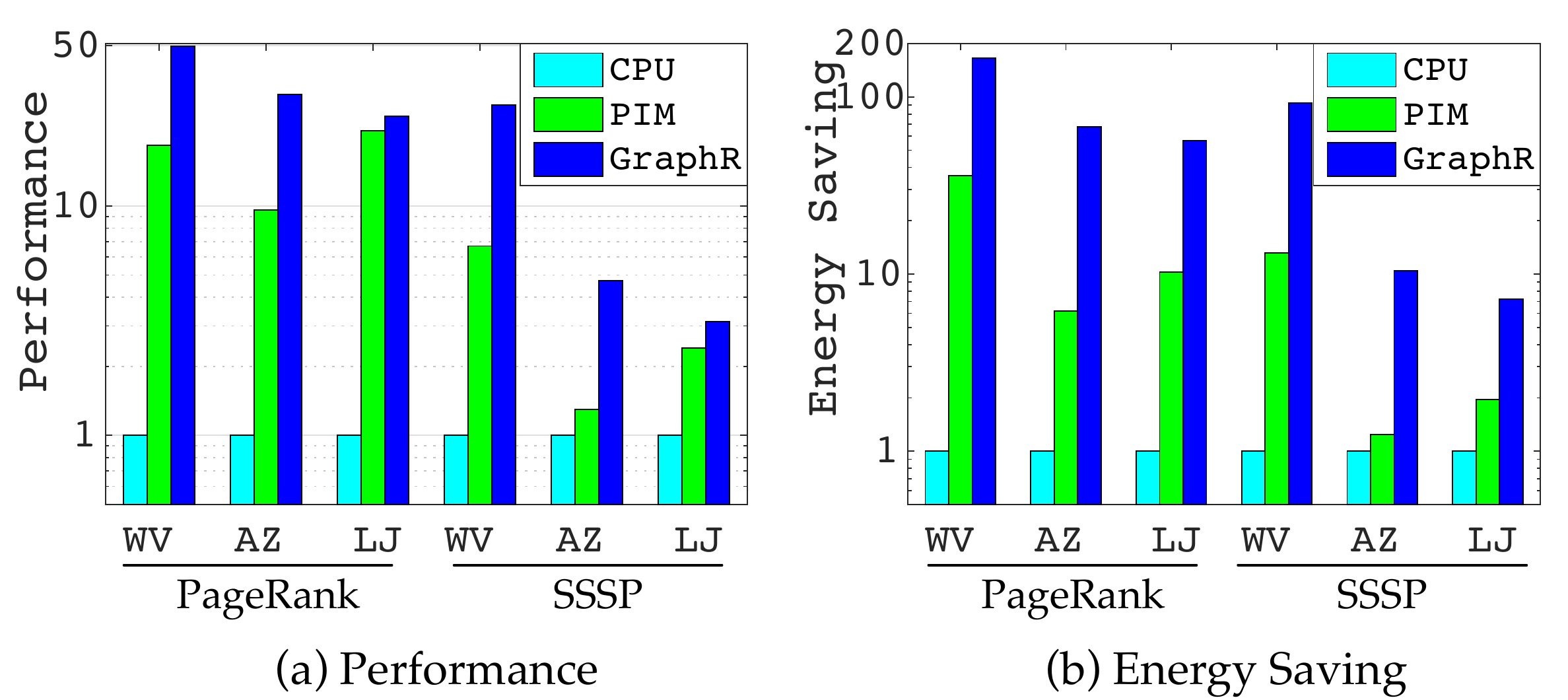}
\vspace{-10pt}
\caption{\scheme\ (a) Performance and (b) Energy Saving Compared to PIM Platform}
\label{topim}
\vspace{-10pt}
\end{figure}

We compare \scheme with a PIM-based architecture (i.e., Tesseract~\cite{ahn2015scalable}).
The performance and energy saving normalized to CPU are shown in Figure~\ref{topim} (a) and (b). \scheme\ gains a speedup of 1.16$\times$ to 4.12$\times$, and 
is 3.67$\times$ to 10.96$\times$ more energy efficiency compared to PIM-based architecture.

\subsection{Sensitivity to Sparsity}

\begin{figure}[htb]
\vspace{-12pt}
\centering
\includegraphics[width=0.98\columnwidth]{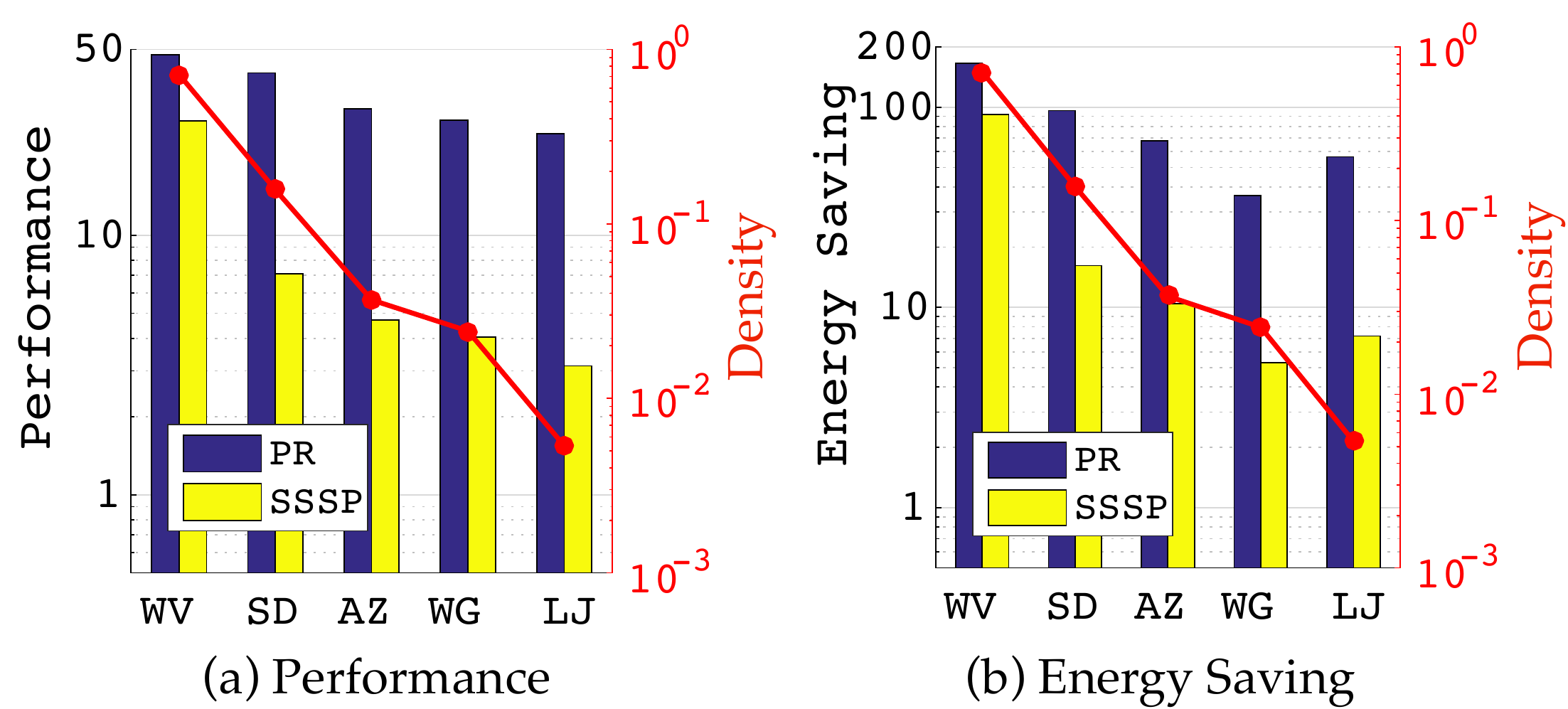}
\vspace{-10pt}
\caption{\scheme\ (a) Performance and (b) Energy Saving with Dataset Density}
\label{sense_sparsity}
\vspace{-10pt}
\end{figure}

We use \#Edge$/$(\#Vertex)$^2$ to represent the density of a dataset, and with the density decreases, the sparsity increases.
Figure~\ref{sense_sparsity} (a) and (b) shows the performance and energy saving of \scheme (compared to the CPU platform) with the density of datasets. With the sparsity increases, the performance and energy saving slightly decreases. Because with the sparsity increases, the number of edge blocks to be traversed will increase, which slows down the edge accessing time and consumes more energy. 

\section{Conclusion}
\label{_conclusion}

This paper presents \scheme, the first ReRAM-based graph processing 
accelerator. 
The key insight of \scheme\ is that if a vertex program of a graph algorithm
can be expressed in sparse matrix vector multiplication (SpMV), it can be 
efficiently performed by ReRAM crossbar.
\scheme\ is a novel accelerator architecture consisting of two components:
{\em memory ReRAM} and {\em graph engine (GE)}. 
The core graph computations are performed in sparse matrix format in GEs (ReRAM crossbars).
With small subgraphs processed by GEs,
the gain of performing parallel operations overshadows
the wastes due to sparsity.
The experiment results show that \scheme\ achieves a 16.01$\times$ (up to 132.67$\times$) speedup and a 33.82$\times$ energy saving on geometric mean compared to a CPU baseline system.
Compared to GPU,
\scheme\ achieves 1.69$\times$ to 2.19$\times$ speedup and
consumes 4.77$\times$ to 8.91$\times$ less energy.
\scheme\ gains a speedup of 1.16$\times$ to 4.12$\times$, and 
is 3.67$\times$ to 10.96$\times$ more energy efficiency compared to PIM-based architecture.

\vspace{3pt}
\noindent\textbf{\large{ACKNOWLEDGEMENT}}
\vspace{0pt}

We thank the anonymous reviewers of HPCA 2018, MICRO 2017 and ISCA 2017 for their constructive and insightful comments. This work was partially supported by NSF-1725456, NSF-1615475, CCF-1717754, CNS-1717984 and DOE-SC0018064.


\newcommand{\BIBdecl}{\setlength{\itemsep}{0.12 em}}
\bibliographystyle{IEEEtranS}
\bibliography{main}

\end{document}